\documentclass[aps,notitlepage,a4paper,superscriptaddress,11pt]{article}

\usepackage[utf8]{inputenc}
\usepackage[dvips]{color}
\usepackage[toc]{appendix}
\usepackage[hang,flushmargin]{footmisc} 
\usepackage{titlesec,titletoc,ntheorem-hyper,authblk,abstract,latexsym,graphicx,microtype,booktabs,amsmath,cleveref,fancyhdr,wrapfig,array,amsfonts, amssymb,geometry,appendix,cite}
\numberwithin{equation}{section}

\newcommand{\R}{\mathbb{R}}
\newcommand{\C}{\mathbb{C}}

\newcommand{\be}{\begin{equation}}
\newcommand{\ee}{\end{equation}}
\newcommand{\bea}{\begin{eqnarray}}
\newcommand{\eea}{\end{eqnarray}}
\newcommand\blfootnote[1]{%
  \begingroup
  
  \addtocounter{footnote}{-1}%
  \endgroup
}

\usepackage{tocloft}    

\usepackage{blindtext,titlefoot}

\title{A pedestrian introduction to coherent and squeezed states}

\author[1]{Bijan Bagchi}
\author[2]{Rupamanjari Ghosh}
\author[3]{Avinash Khare}
\affil[1, 2]{Department of Physics, Shiv Nadar University, Uttar Pradesh 201314, India}
\affil[3]{Department of Physics, Savitribai Phule Pune University,
Pune 411007, India}

\date{\vspace{-6ex}}
\DeclareUnicodeCharacter{2212}{-}
\DeclareUnicodeCharacter{00A8}{ }

\begin{document}
	
\maketitle

 \vspace{2 cm}
	
\begin{abstract}

This review is intended for readers who want to have a quick understanding on the theoretical underpinnings of coherent states and squeezed states which are conventionally generated from the prototype harmonic oscillator but not always restricting to it. Noting that the treatments of building up such states have a long history, we collected the important ingredients and reproduced them from a fresh perspective but refrained from delving into detailed derivation of each topic. By no means we claim a comprehensive presentation of the subject but have only tried to re-capture some of the essential results and pointed out their inter-connectivity. 

\footnote{ E-mails: {bbagchi123@gmail.com}, {rupamanjari.ghosh@snu.edu.in}, 
khare@physics.unipune.ac.in}

\vspace{2 cm}



\end{abstract}

\newpage

\setcounter{tocdepth}{1}
\tableofcontents

\newpage

\section{Introduction}

The story of coherent states dates back to  a paper of Schr\"odinger in $1926$ \cite {Sch} in which he provided a new insight into the
underpinnings of quantum mechanics by constructing the so-called minimum uncertainty wave packets for the harmonic oscillator (HO) potential. The whole exercise depicted a classical motion of a particle moving undistorted in shape as time progressed. All coherent states conform to minimum uncertainty states, the time dependence not hampering the coherence although the underlying parameter could change with it. In the literature three different perspectives \cite{Mmn1, Mmn2, Mmn3} towards construction of coherent states have been followed. These are (i) the ladder-operator method, a specific case in which coherent states emerge as the eigenstates of the annihilation operator; (ii)  displacement-operator method where the vacuum is translated to have a finite excitation amplitude and (iii) employment of minimum uncertainty condition, in which one refers to the  minimization condition of the  Heisenberg uncertainty relation concerning the product of the uncertainties of the canonical position and momentum assuming the value of $\frac{\hbar}{2}$. In practice one can produce a multi-mode coherent state of light from a classical oscillating current in free space. Apart from describing super-fluids and super-conductors, coherent states have relevance to the quantum state of a laser \cite{Walls}.

Coherent states have been in the news off and on since the 1950s (see \cite{Kla1, Kla2} for books covering progress during this period) but the essential breakthrough came in the early sixties through the pioneering works of Glauber\cite{Gla}, and Sudershan \cite{Sud} in the context of quantum optics. Subsequent works especially those of Nieto \cite{Mmn1, Mmn2, Mmn3, Mmn4} and a few others (see the references in  \cite{Gaz1, Fer}) put the subject on a firm theoretical footing. These include employment of the techniques of group theory\footnote{The Heisenberg-Weyl group is the simplest one but not universal.} to define SU(1, 1) and SU(2) coherent states \cite{Barut, Radc} and more generally to arbitrary class of Lie groups \cite{Per}, extension to quantum groups \cite{Bor, Com},  q-deformation \cite{Que1}, $C_\lambda$-extended oscillator \cite{Que2} and position-dependent mass problems \cite{Ruby} as well.  In a recent venture, it has been algebraically shown \cite{Vyas} that accelerating non-spreading wavepackets in a nonrelativistic free particle system \cite{Ber}, having an Airy-type probability distribution, are basically coherent states.  Further, new understandings of coherent states have emerged in the context of quantum information \cite{Gaz2}, entanglement \cite{San, Buk} and field theory concerning the origin of topologically nontrivial gauge fields \cite{Zha}.  For recent advances we refer to \cite{Ver, Hae} for theses and  \cite{Dey, Ort} for reviews and updates.

In addition to quantum optics and atomic theory, applications of coherent states arise in integral quantizations \cite{Berg}, wavelets and their generalizations \cite{Ali}, loop quantum gravity \cite{Fre}, supersymmetric quantum mechanics (SQM) (see, for example, \cite{Hus1}), entropic gravity \cite{Fring},  parity-time symmetric non-Hermitian systems (see for an early work \cite{Bagchi}), non-commutative algebra \cite{Hus2}, to mention a few.  Coherent states are associated with the quantum many-body states of Bose-Einstein condensates \cite{Che}, electronic states in superconductivity \cite{Skg}, signal processing and image processing \cite{Spe}. The idea of pair coherent states has also been considered as a correlated two-mode field \cite{Gsa}. 

It is also worth pointing out that the subject of coherent states has a wide range of current interests that could prove to be worthwhile for the future generation of researchers to pursue. The important emerging fields in this direction point to recent developments in quantum gravity, that make use of notions rooted in geometry, the inherent ideas being exchanged from quantum optics. Other practical use of coherent states concerns quantum communications that include quantum key distribution, quantum digital signatures and quantum fingerprinting \cite{Dim}.

Squeezed states\footnote{The term squeezing was coined by Hollenhorst \cite{Hol}.}, on the other hand, are special states for which one encounters less zero-point fluctuations for one variable while at the same time there is a compensatory increase in fluctuation of the complementary counterpart such that the uncertainty still has the value of $\frac{\hbar}{2}$. Thus, while for a coherent state the position and momentum fluctuations corresponding to the HO tend to be equal in units of $\omega = m =1$, in the case of squeezing, such uncertainties are unequal and less than the value of
$\Delta x$ or $\Delta p$ of the coherent state. However, their  product has the value of $\frac{\hbar}{2}$. Note that the estimate of each uncertainty turns out be $\sqrt{\frac{\hbar}{2}}$ in units of $\omega = m = 1$.

 Geometrically all this means is that while the uncertainty of  the coherent state can be traced by a circle in the phase space, in the case of squeezing the shape is compressed or squeezed to that of an ellipse bearing the same area as the circle and having axes whose lengths are proportional to the radius of the circle depending on the value of the squeezing parameter. Thus an ellipse is the true depiction of a squeezed vacuum state \cite{Lou, Sinha}.  

There is no unique definition that one can ascribe to the notion of a squeezed state \cite{Gsr, Koro}. In its simplest manifestation it is a single-mode of vibration of the electromagnetic field \cite{Whe} and more general in character than a coherent state. Its motion in terms of a single oscillator has been studied \cite{Eke} by introducing quadrature operators and quantum quasi-probability distributions. The application of single-mode squeezing lies in the precision measurements of distances which are done through interferometry.
Another potential application of the vacuum squeezed state is in reducing the quantum noise in gravitational wave detectors
which currently limits their performance in much of the detection band \cite{Bars}. New generation of gravitational wave detectors are known to be GEO 600 in Europe and LIGO in the United States \cite{Lvo}.  In fact new field of gravitational-wave astronomy is likely to be influenced by squeezed light (see, for a review, \cite{Rsc}). A general criterion for squeezing has been noted in \cite{Bar}. For a discussion on some of the nonclassical properties of squeezed states see, for example, \cite{Roy}.

Although early works on squeezing concentrated mostly on the HO potential, generalizations to more complicated situations have been envisaged. Over the years squeezed states have found relevance in several branches of physics, such as the nonlinear optics \cite{Boy}, cavity-QED \cite{Wal}, photonic crystals \cite{Pag}, conformal symmetry \cite{Mply} etc., while applications have been sought in gravitational wave detection where squeezing can reduce noise \cite{Aba, Aas}, quantum images \cite{Lug}, generation of Einstein-Podolsky-Rosen pairs \cite{Gat}, spin squeezing, where the sensitivity of a state is characterized by the class of SU(2) rotations, and which plays a significant role in entanglement detection and high-precision metrology. \cite{Jma},  etc. More recently,

There could be two-mode or multi-mode squeezing as well which are generalizations of the concept of a single-mode squeezed state \cite{Arv}.
Experiments favor the production of two-mode squeezing over a single oscillation mode. In fact, when we are looking at the production of squeezed light, what we observe is a two-mode squeezed light rather than a single oscillation mode. One does not find squeezing in the fluctuation of individual oscillators \cite{Eke}. An example of two-mode squeezed mode laser to use in experiments to provide Wigner function description of quantum nonlocality has also been advanced \cite{VG}. In two-mode squeezing the aspect of noise is present in the pure state, the squeezing parameter plays the role of an effective temperature. Two-mode squeezing can be looked upon as some kind of thermofield state \cite{Kir, Cha} whose evolution is through the Wigner function \cite{Wig}. It is known \cite{Mil} that multi-mode squeezing could result \cite{Bab} from the product of component single-modes, each of which is generated from the ground Fock space. This framework proves rather convenient to handle squeezed multi-mode wavefunction in terms of the normal modes. It also provides a clue to the calculation of Wigner function from the form of single-mode squeezed state defined for the normal modes. On the other hand, for the two-photon squeezed laser-like oscillator \cite{Aga}, single photon decay could cause violation of the conservation law. In this connection see \cite{Gho1} which also gives a detailed account of the theory of a two-photon nonlinear laser with squeezed output. A subsequent work has dealt with the possibility when the lasing medium consists of long-lived atoms \cite{AG}.

Multi-mode squeezed states can be obtained \cite{Mil} by a direct application of a unitary operator on the ground Fock state, its generator being an element of the Cartan subalgebra of $Sp(2n; \mathbb{R})$. However, in this review, we will adopt a different strategy \cite{Bab} to establish a similar idea by concentrating on the two-mode squeezing focusing on different aspects of its theoretical status. Two-mode squeezing has also been interpreted as a system of a pair of coupled oscillators. In fact its generation from a highly correlated state of two oscillators has been demonstrated \cite{Eke}.

The applicability of two-mode squeezing covers a wide range of topics such as quantum optics \cite{Walls}, quantum information \cite{Ade, Brau}, cosmological entropy production \cite{Gasp} and laser interferometers \cite{Schn}, For a recent review we refer to \cite{Gar}. By undertaking squeezing so that the width in the amplitude direction is reduced causing what is termed as "amplitude noise", leads to the corresponding increase in the phase uncertainty. Conversely for an increased amplitude fluctuations the phase squeezed light is decreased. For the case of light, occurrence of squeezing reveals non-classical properties\footnote{For an early work on the observation of non-classical effects in the interference of two photons see \cite{Gho2}. A theoretical treatment of interference of two photons produced via spontaneous parametric down-conversion  show interference effects when fourth-order effects are taken into account \cite{GHOM}.} like non-Poissonian photon statistics and anti-bunching  whose study has proved to be a rather fertile ground for research in quantum optics \cite{Scu}. Such states are even viable even from the experiment point of view employing the use of optical conjugation
\cite{Tei}.


Since most of the basic works on coherent states and squeezed states start from the HO oscillator we highlight and point out in section 2 some of the inherent properties of such a system. In section 3 we review the HO coherent states and touch upon their various properties. In section 4 we discuss pair coherent states. In section 5 we examine the time evolution of coherent states and remark on the classical behavior by deriving the equation of the classical oscillator. In section 6 we address the phase operator and introduce a new set of operators to write down Perelomov coherent states in terms of the generators of the su(1, 1) algebra.
In sections 7 and 8 we consider several aspects of squeezing that include both single-mode and two-mode manifestations. In section 9 we provide a general quantum condition taking care of Bogoliubov invariance. In section 10 we derive a class of coherent states and comment on the generation of squeezed states pertaining to the situation when two potentials are isospectral in the purview of supersymmetric quantum mechanics (SQM).  We also comment on the possibility of deriving coherent states for the specific case of the parity-time symmetric oscillator.  Finally in section 11 we present a brief summary.

\section{A quick look at the harmonic oscillator}

Before we get down to the nitty-gritty of coherent states 
let us recall some of the fundamental relations of the quantum HO\cite{Per, Kla3}. As is well known, a one-dimensional HO for a particle of mass $m$ and angular frequency $\omega$ is characterized by the Hamiltonian

\begin{equation}
H = \frac{\hat{p}^2}{2m} + \frac{1}{2} m \omega^2 \hat{x}^2
\end{equation}
where we adopt the "hat" notation for the position $(\hat{x})$ and momentum $(\hat{p})$ operators. These operators are subject to the quantum condition from which the Heisenberg's uncertainty relation follows

\begin{equation}
[\hat{x},\hat{p}] = i\hbar \quad \Rightarrow \quad \Delta x \Delta p \geq \frac {\hbar}{2}
\end{equation}
Since $\hat{x}$ and $\hat{p}$ cannot be simultaneously diagonalized, we choose a representation, namely the Schr\"odinger representation, in which $\hat{x}$ is diagonal and $\hat{p}$ is given by a first-order differential operator 

\begin{equation}
\hat{x}= x\P, \quad \hat{p}=-i\hbar \frac{d}{dx}
\end{equation}
In the Dirac notation, the position eigenkets obey\cite{Dir1, Gas, Bra, Sak, Acd, Kla4}

\begin{equation}
\hat{x}|x \rangle=x|x \rangle, \quad \langle x|x' \rangle =\delta (x-x'), \quad \int dx |x \rangle \langle x| =\P
\end{equation}
and a similar set of relations for the momentum eigenkets

\begin{equation}
\hat{p}|p \rangle = p|p \rangle, \quad \langle  p|p' \rangle =\delta (p-p'), \quad \int dp |p \rangle \langle p| =\P
\end{equation}

We also note the following forms of the  overlaps 

\begin{equation}
\langle x|p \rangle = \frac{1}{\sqrt{2\pi\hbar}}e^{ipx}, \quad \langle p|x \rangle = \frac{1}{\sqrt{2\pi\hbar}}e^{-ipx}
\end{equation}

Explicitly, the coordinate representation of the time-independent Schr\"odinger equation reads

\begin{equation}
[-\frac{\hbar^2}{2m} +V(x)]\psi_k(x) = E\psi_k(x), \quad k=0,1,2,...
\end{equation}
where, in the context of the HO, $V(x)= \frac{1}{2}m\omega^2 x^2$. The general expression of the wavefunctions $\psi_k (x)$ for such a potential reads (see, for example, \cite{Bra})

\begin{equation}
\psi_k(x)= (\frac{m\omega}{\pi\hbar})^{\frac{1}{4}} \frac{1}{\sqrt{2^k k!}}H_k(x) e^{-\frac{x^2}{2}}, \quad k=0,1,2,...
\end{equation}\\
where $H_k$'s are the Hermite polynomials. The first few are 

\begin{equation}
H_0=1, \quad H_1=x,\quad H_2= 4x^2 -2, \quad H_3 =8x^3 -12x
\end{equation}

Sometimes it is useful to work with the linear combinations of the coordinate and momentum operators. These define the annihilation and creation operators 

\begin{equation}
a=\frac{1}{\sqrt{2m\hbar \omega}} (m\omega x +ip), \quad a^\dagger =\frac{1}{\sqrt{2m\hbar \omega}} (m\omega x -ip)
\end{equation}\\
respectively, on effecting the following set of transformations 

\begin{equation}
\hat{x}=\sqrt{\frac{\hbar}{2m\omega}}(a+a^\dagger), \quad \hat{p}= -i\sqrt{\frac{m\hbar\omega}{2}}(a-a^\dagger)
\end{equation}\\
In terms of $a$ and $a^\dagger$ the quantum condition $(2.2)$ translates to 

\begin{equation}
[a,a^\dagger]=1
\end{equation}
The vacuum is defined by the state $|0\rangle $ which is annihilated by the operator $a$ 
 
\begin{equation}
a|0\rangle  =0
\end{equation}
and the Hamiltonian of the HO takes the form

\begin{equation}
H \equiv \frac{\hbar\omega}{2}[a,a^\dagger]_+=\hbar\omega(a^\dagger a +\frac{1}{2})
\end{equation}

On the other hand, in the Fock representation, the $n$-th eigenstate, $n=0,1,2,...$, is obtained by repeated application of the creation operator $a^\dagger$ on $|0 \rangle$ 

\begin{equation}
|n \rangle = \frac{(a^\dagger)^n}{\sqrt{n!}} |0 \rangle
\end{equation}
Furthermore, the followings results 

\begin{equation}
a|n \rangle = \sqrt{n}|n-1 \rangle, \quad a^\dagger|n \rangle = \sqrt{n+1} |n+1 \rangle
\end{equation}
show that the annihilation operator lowers the number of particles in a given state by one unit while the creation operator raises it by one unit as well. The creation operator is the adjoint of the annihilation operator and vice versa.\\

Along with the Hamiltonian $H$ one can define a number operator $N$ 

\begin{equation}
N= a^\dagger a
\end{equation}
having the property 

\begin{equation}
N|n \rangle = n|n \rangle, \quad n=0,1,2,...
\end{equation}
where the eigenvalues correspond to the set of natural numbers which also includes the null value. The number operator $N$ satisfies the commutation relations 

\begin{equation}
[N,a]=-a, \quad [N,a^\dagger] =a^\dagger
\end{equation}

Finally, it is easy to see that the expectation of $x$ in the $n$-th state vanishes

\begin{equation}
\langle \hat{x}\rangle_n= \langle n|\hat{x}|n \rangle =\sqrt{\frac{\hbar}{2m\omega}} \langle n|(a+a^\dagger)|n \rangle =0
\end{equation}
where we used the property that $a$ or $a^\dagger$ has zero matrix element between identical states. An analogous result for the momentum state $p$ holds

\begin{equation}
\langle \hat{p} \rangle_n= \langle n|\hat{p}|n \rangle =-i\sqrt{\frac{m\hbar\omega}{2}} \langle n|a-a^\dagger|n \rangle =0
\end{equation}
   
This concludes our discussion on the HO in which we summarized its essential properties that we will need for later treatment of coherent and squeezed states.

\section{HO coherent states}

Taking cue from where we just left we observe that since

\begin{equation}
\langle 0|(a+a^\dagger)^2|0 \rangle = \langle 0|(a+a^\dagger)(a+a^\dagger)|0 \rangle = \langle 0|aa^\dagger|0 \rangle = 1
\end{equation}\\
and similarly

\begin{equation}
\langle 0|(a-a^\dagger)^2|0 \rangle = \langle 0|(a-a^\dagger)(a-a^\dagger)|0 \rangle =-\langle 0|aa^\dagger|0\rangle = -1
\end{equation}\\
where we used $\langle i|j\rangle =\delta_{ij}$ and $\langle 0|a^\dagger a|0\rangle =0$, it follows from $(2.11), (3.1)$ and $(3.2)$ that the product of the vacuum values of $\hat{x}^2$ and $\hat{p}^2$ is 

\begin{equation}
\langle x^2 \rangle_0 \langle p^2 \rangle_0 = \frac{\hbar^2}{4}
\end{equation}\\
showing that the vacuum $|0\rangle $ satisfies the minimal uncertainty condition

\begin{equation}
(\Delta x)^2 (\Delta p)^2 = \frac{\hbar^2}{4}
\end{equation}\\
where we used the notation $\Delta O=O-\langle O \rangle $, with $\langle O \rangle = \langle i|O|i \rangle$ corresponding to the expectation in the state vector $|i \rangle$.

This feature, however, is not shared by any arbitrary state $|n \rangle$  of the HO. For a demonstration we consider the following expectation values 

\begin{equation}
\langle n|(a+a^\dagger)^2|n \rangle= \langle n|(a+a^\dagger)(a+a^\dagger)|n \rangle= \langle 0|2a^\dagger a +[a,a^\dagger]|0 \rangle = 2n+1
\end{equation}\\
and 

\begin{equation}
\langle n|(a-a^\dagger)^2|n \rangle= \langle n|(a-a^\dagger)(a-a^\dagger)|n \rangle = - \langle 0|2a^\dagger a +[a,a^\dagger]|0 \rangle = -(2n+1)
\end{equation}\\
where we used $(2.12)$ and $(2.18)$. In other words, the product $\langle x^2 \rangle_n \langle p^2 \rangle_n$ becomes

\begin{equation}
\langle x^2 \rangle_n \langle p^2 \rangle_n = \frac{\hbar^2}{4} (2n+1)^2
\end{equation}\\
which implies 

\begin{equation}
(\Delta x)^2 (\Delta p)^2 = \frac{\hbar^2}{4}(2n+1)^2, \quad n=0,1,2,...
\end{equation}\\
Comparing with $(2.2)$  it is clear that the minimum value holds only for $n=0$.\\

A natural question to ask \cite{Iwa} is whether the vacuum is the only minimal uncertainty state. The
answer, as we shall presently see, is in the negative. Towards this end, suppose that there exists  a state $|\alpha\rangle $ which is an eigenstate of the annihilation operator $a$

\begin{equation}
a|\alpha \rangle = \alpha |\alpha\rangle, \quad \alpha \in \C
\end{equation}
As such

\begin{equation}
\langle \alpha|a^\dagger a|\alpha \rangle =|\alpha|^2
\end{equation}\\
We also have the following set of relations

\begin{equation}
\langle \alpha|(a+a^\dagger)(a+a^\dagger)|\alpha \rangle = (\alpha +\alpha^*)^2 +1, \quad \langle \alpha|(a-a^\dagger)(a-a^\dagger)|\alpha \rangle = (\alpha -\alpha^*)^2 -1
\end{equation}\\
resulting in the following representation of the uncertainty $\langle x\rangle _\alpha$ 

\begin{equation}
\langle x \rangle_\alpha = \sqrt{\frac{\hbar}{2m\omega}} \langle a+a^\dagger \rangle_\alpha =\sqrt{\frac{\hbar}{2m\omega}} (\alpha +\alpha^*)
\end{equation}
where the right side is a real number. Further

\begin{equation}
\langle x^2 \rangle_\alpha = \frac{\hbar}{2m\omega} \langle (a+a^\dagger)^2 \rangle_\alpha =\frac{\hbar}{2m\omega} [(\alpha +\alpha^*)^2 +1]
\end{equation}
We therefore obtain

\begin{equation}
\langle (\Delta x)^2 \rangle_\alpha = \langle x^2 \rangle_\alpha - (\langle x \rangle_\alpha)^2 =\frac{\hbar}{2m\omega} 
\end{equation}
Similarly for $\langle (\Delta p)^2\rangle _\alpha$ we get

\begin{equation}
\langle (\Delta p)^2\rangle _\alpha = \langle p^2 \rangle_\alpha - (\langle p \rangle_\alpha)^2 =\frac{\hbar m\omega}{2} 
\end{equation}
Taking the product of $(3.14)$ and $(3.15)$ yields 

\begin{equation}
\langle (\Delta x)^2 \rangle \langle (\Delta p)^2 \rangle = \frac{\hbar^2}{4}
\end{equation}
for the state $|\alpha \rangle$. It signals the same minimal uncertainty value as found for the vacuum state. The $|\alpha \rangle$ 's are termed as the coherent states. We emphasize that every possible choice of $\alpha$ admits of a different coherent state. Coherent states\footnote{Note that angular momentum coherent states analogous to the ones of the HO can also be defined \cite{Dut, Band}.} correspond to the most classical states of the harmonic oscillator. 

That the coherent state cannot conform to any of the basis states $|n \rangle$ of the HO is clear from the observation that a basis state $|n \rangle$, other than the solitary vacuum, cannot be a minimal uncertainty state. We therefore inquire into the following issues:\\

(1) How to relate the coherent state $|\alpha \rangle$ with the basis states $|n \rangle $ of the harmonic oscillator?\\

(2) How to generate the coherent state $|\alpha \rangle$ from the vacuum $|0 \rangle$ ?\\

To answer the first question, let us express $|\alpha \rangle$ as a linear combination of the states $|n\rangle $

\begin{equation}
|\alpha \rangle = \sum_{n=0}^\infty c_n|n \rangle
\end{equation}\\
Inserting $\sum_n |n \rangle \langle n| =1$ to re-cast it 

\begin{equation}
|\alpha \rangle = \sum_{n=0}^\infty |n \rangle \langle n|\alpha \rangle
\end{equation}\\
we use $(2.15)$ to write
\begin{equation}
\langle n|\alpha \rangle = \frac{\alpha^n}{\sqrt{n!}} \langle 0|\alpha \rangle
\end{equation}\\
This implies from $(3.18)$ the form  

\begin{equation}
|\alpha \rangle = \langle 0|\alpha \rangle \sum_{n=0}^\infty \frac{\alpha^n}{\sqrt{n!}}|n \rangle
\end{equation}\\
Utilizing the normalization condition $\sum_n \langle \alpha|n \rangle \langle n|\alpha \rangle =1$ then gives  

\begin{equation}
1= |\langle 0|\alpha \rangle|^2 \sum_{n=0}^\infty \frac{|\alpha|^{2n}}{n!}= |\langle 0|\alpha \rangle|^2 e^{|\alpha|^2}
\end{equation}\\
suggesting that the overlap $\langle 0|\alpha \rangle$ is given by
\begin{equation}
\langle 0|\alpha \rangle = e^{-\frac{1}{2}|\alpha|^2}
\end{equation}\\
which holds up to a phase factor. Hence the final form of the coherent state $|\alpha \rangle$ is 

\begin{equation}
|\alpha \rangle = e^{-\frac{1}{2}|\alpha|^2}\sum_{n=0}^\infty \frac{\alpha^{n}}{\sqrt{n!}} |n \rangle
\end{equation}\\
The value $\alpha =0$ is consistent with the vacuum $|0 \rangle$ while the amplitude of finding $|\alpha \rangle$ in the state $|n \rangle$ is 

\begin{equation}
\langle n|\alpha \rangle= e^{-\frac{1}{2}|\alpha|^2}\left ( \frac{\alpha^{n}}{\sqrt{n!}} \right )
\end{equation}\\

With $\langle n\rangle  = |\alpha|^2$,  it follows that

\begin{equation}
|\langle n|\alpha \rangle|^2 = \frac{\langle n \rangle^n}{n!} \exp (-\langle n \rangle)
\end{equation}
signalling a Poisson distribution for $n$ photons to be in the coherent state. 

That the coherent states are not orthogonal can be noticed by taking the scalar product of two such states $|\alpha \rangle$ and $|\alpha' \rangle$ which shows

\begin{eqnarray}
&&\langle \alpha|\alpha' \rangle = e^{-\frac{1}{2}|\alpha|^2} e^{-\frac{1}{2}|\alpha'|^2} \sum_{m=0}^\infty \sum_{n=0}^\infty \frac{\alpha^{* m}}{\sqrt{m!}} \frac{\alpha'^{n}}{\sqrt{n!}} \nonumber\\ 
&&= e^{(-\frac{1}{2}|\alpha|^2 - \frac{1}{2}|\alpha'|^2)}e^{\alpha^* \alpha'} \nonumber\\
&& = e^{-|\alpha - \alpha'|^2}
\end{eqnarray}
Thus, unless $|\alpha - \alpha'| >> 1$ when the states $|\alpha \rangle$ and $|\alpha' \rangle$ are approximately orthogonal, $|\alpha \rangle$ and $|\alpha' \rangle$ are generally non-orthogonal.

In effect this means that if we write $|\alpha'\rangle$ as

\begin{equation}
|\alpha' \rangle = \frac{1}{\pi} \int d^3 \alpha |\alpha \rangle \langle\alpha|\alpha' \rangle
\end{equation}\\
and insert for $\langle \alpha|\alpha'\rangle $ the just derived result one clearly sees that the coherent states are not linearly independent but in fact overcomplete..

One can prove the completeness result by expressing

\begin{equation}
\frac{1}{\pi} \int d^3 \alpha |\alpha \rangle \langle \alpha|= \frac{1}{\pi} \sum_{m=0}^\infty \sum_{n=0}^\infty \frac{1}{\sqrt{m!}} \frac{1}{\sqrt{n!}} |n \rangle \langle m| \int d^2\alpha e^{-|\alpha|^2} \alpha^n (\alpha^*)^m
\end{equation}\\
Setting $\alpha = |\alpha| e^{i\phi}$ and noting that $\int_0^{2\pi} d\phi e^{i(n-m)\phi} =2\pi \delta_{nm}$ it follows that

\begin{equation}
 \int d^2\alpha e^{-|\alpha|^2} \alpha^n (\alpha^*)^m = 2\pi \int_0^\infty dr r e^{-r^2} r^{2n}
\end{equation}\\
The integral in the right side on substituting $r^2 =z$ turns out to be of the form $\frac{1}{2} \Gamma (n+1) =\frac{1}{2} n!$ and furnishes the completeness relation

\begin{equation}
\frac{1}{\pi} \int d^3 \alpha |\alpha \rangle \langle \alpha|= \sum_n |n \rangle \langle n| = \P
\end{equation}\\
For an operator representation of the coherent states we need to make use of the well known  Baker-Campbell-Hausdorff (BCH) formula for two operators $A$ and $B$  

\begin{equation}
e^{\lambda A} B e^{-\lambda A} =B +\lambda [A,B]+ \frac{\lambda^2}{2!}[A,[A,B]] + \frac{\lambda^3}{3!}[A,[A,[A,B]]]+,,,,\quad \lambda \in \R
\end{equation}\\
An off-shoot is that if  $[A,B]$ commutes with each of $A$ and $B$ then 

\begin{equation}
e^{A+B} =e^A e^B e^{-\frac{1}{2} [A,B]}
\end{equation}\\
An implication is the result

\begin{equation}
e^{\alpha a^{\dagger} -\alpha^* a}=e^{-\frac{1}{2}|\alpha|^2+\alpha a^\dagger} e^{-\alpha^* a}
\end{equation}\\
where we used $(2.12)$. When applied on the vacuum one finds 

\begin{equation}
e^{\alpha a^{\dagger} -\alpha^* a}|0 \rangle =e^{-\frac{1}{2}|\alpha|^2+\alpha a^\dagger}|0 \rangle
\end{equation}\\
because by $(2.13)$ we can write $e^{-\alpha^* a}|0\rangle  =0$.

Since we can express 

\begin{equation}
\sum_{n=0}^\infty \frac{\alpha^{n}}{\sqrt{n!}} |n \rangle =\sum_{n=0}^\infty \frac{\alpha^{n}}{\sqrt{n!}} (a^\dagger)^n |0 \rangle =e^{\alpha a^\dagger}|0 \rangle
\end{equation}\\
it gives, using $(3.23)$ and $(3.33)$, another form for $|\alpha \rangle$

\begin{equation}
|\alpha \rangle= e^{-\frac{1}{2}|\alpha|^2+\alpha a^\dagger}|0 \rangle =e^{\alpha a^{\dagger} -\alpha^* a}|0 \rangle
\end{equation}\\
The above representation is frequently used in the theory of coherent states.

To extract the coordinate representation of the coherent state, we substitute the definition of the $a^\dagger$ operator in the above representation. Indeed using $(2.10)$ we find  

\begin{equation}
\langle y|\alpha \rangle =  e^{-\frac{1}{2}|\alpha|^2} \langle y|e^{\alpha \sqrt{\frac{m\omega}{2\hbar}}(x-\frac{ip}{m\omega})}|0 \rangle
\end{equation}
which can be transformed to 

\begin{equation}
\langle y|\alpha \rangle =  e^{-\frac{1}{2}|\alpha|^2}e^{\alpha \sqrt{\frac{m\omega}{2\hbar}}(y-\frac{\hbar}{m\omega}\frac{d}{dy})} \langle y|0 \rangle
\end{equation}
Since from $(2.8)$ and $(2.9)$ the ground state wave function of the HO reads (in the units of $\hbar = \omega = m = 1)$

\begin{equation}
\psi_0(x) \equiv \langle x|0 \rangle = (\frac{m\omega}{\pi\hbar})^\frac{1}{4}e^{-\frac{1}{2}x^2}
\end{equation}
$\langle y|\alpha\rangle $ assumes the form 

\begin{equation}
\langle y|\alpha\rangle  = (\frac{m\omega}{\pi\hbar})^\frac{1}{4} e^{-\frac{1}{2}|\alpha|^2}e^{\frac{\alpha}{\sqrt{2}}
(y-\frac{d}{dy})}e^{-\frac{1}{2} x^2}
\end{equation}
We are thus led to the following form of the coherent state \cite{Dod}

\begin{equation}
\psi_c (x) \equiv \langle x|\alpha \rangle = (\pi)^{-\frac{1}{4}} \exp \left [-\frac{1}{2} x^2 + \sqrt{2} \alpha x - \frac{1}{2} \alpha^2 - \frac{1}{2} |\alpha|^2 \right ], \quad \alpha \in \mathbb{C}
\end{equation}
where we have adopted the notations $\langle x\rangle  = \sqrt{2} \alpha_R$ and $\langle p\rangle  = \sqrt{2} \alpha_I$, $\alpha_R$ and $\alpha_I$ being the real and imaginary components of $\alpha$. Such a wave function minimizes the Heisenberg uncertainty relation (2.2). It also transpires that $\psi (x)$ is a Gaussian and depicts a similar feature as the ground state of the harmonic oscillator but with a shifted position. It also possesses a phase which from $(3.41)$ can be seen to be proportional to the position $x$. Further, within the confines of the HO potential, the width of the wave function $\psi (x)$ is that of its ground state. \\

We now turn to the second question.\\

The coherent state can also be generated from the vacuum by applying the displacement operator on $|0\rangle $. The displacement operator $D$ is defined by 

\begin{equation}
D(\alpha) = e^{\alpha a^{\dagger} -\alpha^* a}=e^{-\frac{1}{2}|\alpha|^2} e^{\alpha a^\dagger} e^{-\alpha^* a}= e^{\frac{1}{2}|\alpha|^2} e^{-\alpha^* a} e^{\alpha a^\dagger}
\end{equation}\\
where we employed $(3.33)$.  It has a unitary complex feature. Two important properties of the displacement operator are 

\begin{eqnarray}
&& D^\dagger (\alpha) a D(\alpha) = a +\alpha, \\
&& D(\alpha +\beta) = e^{\frac{1}{2} (\alpha \beta^* - \alpha^* \beta)} D(\alpha) D(\beta)    
\end{eqnarray}
While the first relation speaks of producing a displacement, the second one says that the result of two successive displacements is equivalent to another displacement except that one has to take into account the appearence of a phase factor. 

In addition to $(3.43)$ and $(3.44)$ the following results are also valid

\begin{eqnarray}
&& D^\dagger (\alpha) = D^{-1} (\alpha) =D(-\alpha) \\
&& D^\dagger (\alpha) a^\dagger D(\alpha) = a^\dagger +\alpha^*
\end{eqnarray}

Since $|\alpha\rangle  =D(\alpha)|0\rangle $, it means that the displacement operator can act on the vacuum to produce the coherent state. 

A few words on the group theoretical representation of the Perelomov coherent states. Defining the displacement operator as (see, for example, \cite{Oje}) 

\begin{eqnarray}
D(\xi) = e^{\xi K^+ -\xi^* K^-}, \quad D(-\xi) = D^\dagger (\xi)
\end{eqnarray}
where $\xi \in \mathbb{C}$ and hence can be put in the form $\xi = -\frac{1}{2} r e^{-i\phi}$ with $ -\infty < r < \infty$ and $0 \leq \phi \leq 2\pi$.  The operators $K^+, K^-$ along with $K^0$ satisfy the algebra of $SU(1, 1)$ group namely, 

\begin{eqnarray}
[K^\pm, K^0] = \mp K^\pm, \quad [K^+, K^-] = - 2K^0
\end{eqnarray}
which in the Fock space operate according to 

\begin{eqnarray}
&& K^+ |k, n\rangle  = \sqrt{(n+1)(2k +n)} |k, n+1\rangle  \\
&&  K^- |k, n\rangle  = \sqrt{n(2k +n-1)} |k, n-1\rangle  \\
&&  K^0 |k, n\rangle  = (k+n) |k, n\rangle 
\end{eqnarray}
corresponding to the states $|k, n\rangle $ where $n=0,1,2,...$ having $|k, 0\rangle $ as the lowest normalized state. The Casimir operator for such a set of generators is given by $K^2 = K_0^2 - \frac{1}{2} (K^+K^- + K^- K^+)$ and obeys $K^2 = k(k-1)$ for any irreducible representation. 

The displacement operator $D(\xi)$ can be disentangled to read as the product of the exponentials \cite{Gerr}

\begin{eqnarray}
D(\xi) = e^{\kappa  K^+} e^{\rho  K^0} e^{-\kappa^*  K^-}
\end{eqnarray}
where $\kappa = -\tanh (\frac{1}{2} r) e^{-i\phi}$ and $\rho = \ln (1-|\kappa|^2)$.

The general form of the coherent state can be expressed as 

\begin{eqnarray}
D(\kappa) = D(\xi) |k, 0 \rangle =  (1-|\kappa|^2)^k \sum_{j=0}^\infty \sqrt{\frac{\Gamma (n+2k)}{j! \Gamma (2k)}} \kappa^j |k, j \rangle
\end{eqnarray}

\section{Time evolution and classical behaviour of coherent states}

In the Schr\"odinger picture the state vector evolves with time but the Hamiltonian does not depend on time. The fundamental governing equation is 

\begin{equation}
i\hbar \frac{\partial U}{\partial t} =HU(t)
\end{equation}\\
where $U$ is the evolution operator $|\psi(t) \rangle = U(t)|\psi(0) \rangle$ whose solution is 

\begin{equation}
U(t) = e^{-\frac{i}{\hbar}Ht}
\end{equation}\\ 
For the time-developed coherent state $|\alpha\rangle $ we write 

\begin{equation}
|\alpha, t \rangle = e^{-\frac{i}{\hbar}Ht}|\alpha (0) \rangle 
\end{equation}\\
where $|\alpha, t\rangle $ is the coherent state. 

Employing $(3.24)$ we cast the above equation to the form

\begin{equation}
|\alpha(t) \rangle = U(t)|\alpha \rangle = e^{-\frac{i}{\hbar}Ht}|\alpha \rangle =
e^{-\frac{1}{2}|\alpha|^2}\sum_{n=0}^\infty \frac{\alpha^{n}}{\sqrt{n!}}e^{-\frac{iHt}{\hbar}} |n \rangle
\end{equation}\\
where $\alpha = \alpha (0)$. 

Using the HO Hamiltonian $(2.14)$, $|\alpha, t\rangle  $ turns out to be 

\begin{equation}
|\alpha, t \rangle = U(t)|\alpha \rangle = e^{-\frac{i}{\hbar}Ht}|\alpha \rangle=
e^{-\frac{1}{2}|\alpha|^2}\sum_{n=0}^\infty \frac{\alpha^{n}}{\sqrt{n!}}e^{-i\omega t(n+\frac{1}{2})}|n \rangle
\end{equation}\\
A little calculation gives

\begin{equation}
|\alpha, t \rangle = e^{-\frac{i}{2} \omega t} |e^{-i\omega t} \alpha (0) \rangle = |\alpha (t)\rangle
\end{equation}\\
implying that the character of the coherent state does not change under time evolution.

To calculate $\langle x \rangle$ in the state $|\alpha (t) \rangle$ we readily see 

\begin{equation}
\langle\alpha (t)|\hat{x}|\alpha (t) \rangle =\sqrt{\frac{\hbar}{2m\omega}}
e^{-|\alpha|^2}\sum_{m=0}^\infty \sum_{n=0}^\infty \langle m|a+a^\dagger|n \rangle \frac{\alpha^{*m}}{\sqrt{m!}}\frac{\alpha^{n}}{\sqrt{n!}}e^{-i\omega t(n-m)}
\end{equation}\\
Since by $(2.16)$

\begin{equation}
\langle m|a|n \rangle = \sqrt{n} \delta_{m,n-1}, \quad \langle m|a^\dagger|n \rangle  = \sqrt{n+1} \delta_{m,n+1}
\end{equation}\\ 
we can reset 

\begin{equation}
\langle \alpha (t)|\hat{x}|\alpha (t) \rangle =\sqrt{\frac{\hbar}{2m\omega}}
e^{-|\alpha|^2}\sum_{m=0}^\infty \sum_{n=1}^\infty \frac{\alpha^{*m}}{\sqrt{m!}}\frac{\alpha^{n}}{\sqrt{n!}}(\sqrt{n} \delta_{m,n-1} + \sqrt{n+1} \delta_{m,n+1} ) e^{-i\omega t(n-m)}
\end{equation}\\

which simplifies to

\begin{equation}
\langle \alpha (t)|\hat{x}|\alpha (t) \rangle =\sqrt{\frac{\hbar}{2m\omega}}
e^{-|\alpha|^2} \left (\sum_{n=1}^\infty \frac{|\alpha|^{2(n-1)}}{(n-1)!}\alpha e^{-i\omega t}+ \sum_{n=1}^\infty \frac{|\alpha|^{2n}}{(n)!}\alpha^* e^{i\omega t} \right )
\end{equation}\\
Putting $\alpha =|\alpha|e^{i\phi}$, where $\phi$ is a real number, we arrive at

\begin{equation}
\langle \alpha (t)|\hat{x}|\alpha (t) \rangle =\sqrt{\frac{\hbar}{2m\omega}}
e^{-|\alpha|^2}e^{|\alpha|^2} \left (|\alpha|[e^{-i\omega t +\delta} + e^{i\omega t +\phi}] \right ) 
\end{equation}\\
which reduces to

\begin{equation}
\langle \alpha (t)|\hat{x}|\alpha (t) \rangle =2\sqrt{\frac{\hbar}{2m\omega}}
|\alpha|\cos (\omega t -\phi)
\end{equation}\\
It shows an oscillatory behavior of the classical oscillator. In fact if we double differentiate we do get the equation of the simple harmonic motion

\begin{equation}
\langle \ddot{x} \rangle  +\omega^2 \langle x \rangle=0
\end{equation}\\
In other words, coherent states, which are truly quantum states, depict a classical behavior.

\section{Pair coherent states}

Pair coherent states are defined as the simultaneous eigenstates of an operator that annihilates photons in pairs and that of the operator that gives the relative occupation number in the two-modes \cite{Gsa}. Such coherent states exhibit a number of interesting properties including the feature of sub-Poissonian statistics, correlations in the photon number fluctuations in the two-modes and violations of Cauchy-Schwarz inequalities. 

More recently, the subject of entanglement has been studied in pair coherent state\footnote{The wave function of pair coherent states is typically non-Gaussian for a two-mode radiation field. Nonclassical properties of pair coherent state has attracted much attention along with violation of Bell inequalities \cite{Gil, Tara}.} and the inseparability of the pair-coherent state was examined from in the light of the 
non-classicality of the Glauber-Sudarshan P-function. The latter, as is well known, gives
a quasi-probability distribution in phase space, such a distribution can take negative and singular
values for non-classical fields \cite{Gsb}. Further, pair coherent states are of relevance in quantum teleportation \cite{Gag} and quantum information processes \cite{Wu}.  Also there have been  extensions to the generalized pair coherent states where two coherent variables have a certain phase difference (see, for example, \cite{Has}).

If $a$ and $b$ denotes modes of the photon then the pair annihilation operator is defined by  

\begin{equation}
ab |\zeta, q \rangle = \zeta |\zeta, q \rangle, \quad q>0
\end{equation}\\
where $q$ denotes the photon number difference for the the two modes in the field \cite{Bbd}

\begin{equation}
(a^\dagger a - b^\dagger b) |\zeta, q \rangle = q |\zeta, q \rangle
\end{equation}\\
The exponential form of the pair coherent state is \cite{Wang1}

\begin{equation}
|\zeta, q \rangle = \zeta |\zeta, q \rangle, \quad q>0
\end{equation}\\
up to a normalization constant.
The coherent state $(5.1)$ is a specific case of the two-mode nonlinear coherent state 

\begin{equation}
f(N_a, N_b)ab |\alpha, f, q \rangle = \zeta |\alpha, f, q \rangle
\end{equation}\\
where $N_a$ and $N_b$ are number operators which goes over to $(5.1)$ when $f=1$.

In this connection it is worthwhile to mention the two-mode Perelomov coherent state which reads

\begin{equation}
|\xi, q \rangle = e^{(\xi a^\dagger b^\dagger - \xi^* a b)} |q, 0 \rangle
\end{equation}\\
obeying the equation 

\begin{equation}
\frac{2}{2+ q + N_a + N_b}ab |\xi, q \rangle = \frac{\xi \tanh (|\xi|)}{|\xi|} |\xi, q \rangle
\end{equation}\\
showing that the Perelomov coherent states are two-mode nonlinear coherent states, the coefficient in the right side serving as a nonlinear function. In the two-mode Fock space the $|\zeta, q\rangle $ in $(5.5)$ reads

\begin{equation}
|\xi, q \rangle = exp\left [ \frac{\xi \tanh (|\xi|)}{|\xi|} a^\dagger b^\dagger \right ] |q, 0 \rangle
\end{equation}\\
with a suitable normalization constant. 
In the context of photon-added coherent states proposed in \cite{Gst} one can show that these are nonlinear coherent states \cite{Siva}. A generalization was carried out in \cite{Wang2}. We refer to \cite{Sha} for a unified approach to multiphoton coherent states. The techniques formulated there can be generalized to deformed bosons.

About pair coherent states a couple of remarks are in order. In a two-photon medium such coherent states arise due to competition between nonlinear gain and nonlinear absorption where the threshold refers to the condition of linear gain being equal to the linear absorption. In a two-level system the problem of interaction of fields is difficult to tackle analytically unless one considers the situation of the interaction with only one of the correlated modes. 

There is another class of coherent states called the parity-pair (PP) coherent states \cite{Spi}. It has the form 

\begin{equation}
|\zeta, q \rangle_{PP} = \sum_{n=0}^\infty \sqrt{\frac{q!}{n! (n+q)!}} \zeta^n (-1)^{-\frac{n(n-1)}{2}} |n+q, n \rangle
\end{equation}\\
It can be re-expressed as a superposition of two pair coherent states having a phase difference of $\pi$

\begin{equation}
|\zeta, q \rangle_{PP} = \frac{1}{\sqrt{2}} \left ( e^{-i\frac{\pi}{4}} |i\zeta, q\rangle  + e^{i\frac{\pi}{4}} |-i\zeta, q \rangle \right )
\end{equation}\\
It resembles the parity harmonic oscillator coherent states \cite{Yur}.

\section{The phase operator}

Let us introduce \cite{Hei, Dir2} a phase angle $\phi$ to define  $a$ and $a^\dagger$ according to

\begin{equation}
a = \sqrt{N+1} e^{-i\phi}, \quad a^\dagger = e^{i\phi} \sqrt{N+1}
\end{equation}\\
This means that

\begin{equation}
e^{-i\phi} =  \left (\frac{1}{\sqrt{N+1}}\right ) a, \quad e^{i\phi} = a^\dagger \left (\frac{1}{\sqrt{N+1}}\right ) 
\end{equation}\\
While $e^{i\phi} e^{-i\phi} = 1$ follows trivially, it is not difficult to check 

\begin{equation}
e^{-i\phi} e^{i\phi} \neq 1
\end{equation}\\
pointing to the non-Hermiticity of the phase operator. In other words the exponential $e^{i\phi}$ is not a unitary operator.

Since corresponding to the operators $X$ and $Y$ the commutator $[X, Y] = k$ implies $[X, e^{cY}] = ck e^{cY}$, where $c$ and $k$ are real parameters,  it follows that 

\begin{equation}
[N, e^{i\phi}] = e^{\phi}, \quad [N, \phi] =-i
\end{equation}\\
on putting $X=N, Y=\phi$ and $k = 1$. A natural outcome is the  uncertainty constraint between the phase and number operator

\begin{equation}
\Delta \phi \Delta N \geq \frac{1}{2}
\end{equation}\\
It follows that for a known phase an uncertain number of particles is implied. This is a standard feature of coherent states in contrast to the Fock states where because of a definite number of particles the phase is uncertain.

We now turn to an interpretation of the phases $e^\phi$ and $e^{-\phi}$ which are indeed operators because the equations in the right side of $(6.2)$ are so.
Using the notation $\Phi$ for the representation of $\phi$, more specifically $\Gamma^{\pm}$ for $e^{\pm i\Phi}$, we write \cite{Sus}

\begin{equation}
e^{-i\Phi}|n\rangle  \rightarrow \Gamma^- = \left (\frac{1}{\sqrt{N+1}}\right ) a, \quad e^{i\Phi}|n\rangle  \rightarrow \Gamma^+ = a^\dagger \left (\frac{1}{\sqrt{N+1}}\right ) 
\end{equation}\\
the following set of relations immediately follows

\begin{eqnarray}
&& \Gamma^{-} |n \rangle = |n-1 \rangle, \quad n= 1,2,...\nonumber \\
&& \Gamma^{-} |n \rangle = 0, \nonumber \\
&& \Gamma^{+} |n \rangle = |n+1 \rangle, \quad n= 0,1,2,...\nonumber \\
&& \Gamma^{-}\Gamma^{+} |n \rangle = |n \rangle, \quad n= 0,1,2,...\nonumber \\
&& \Gamma^{+}\Gamma^{-} |n \rangle = |n \rangle, \quad n= 1,2,...\nonumber \\
&& \Gamma^{+}\Gamma^{-} |0 \rangle = 0  \label{fac1}
\end{eqnarray}

Identifying \cite{Car1, Car2}

\begin{equation}
\Gamma^{+} + \Gamma ^{-} = 2 \cos\Phi, \quad \Gamma^{+} - \Gamma ^{-} = -2i \cos\Phi
\end{equation}\\
has the implications

\begin{equation}
\left (\Delta \cos\phi \right ) \Delta N \geq \frac{1}{2} \left |\sin\Phi \right |, \quad \left (\Delta \sin\phi \right ) \Delta N \geq \frac{1}{2} \left |\cos\Phi \right |
\end{equation}\\
These can be considered as equivalent to the ones in $(5.5)$.

Interestingly a realization of $SU(1, 1)$, which was discussed in $(3.48) - (3.52)$, is possible \cite{Kli, Suk1} if we introduce additional pair of operators $R^+$ and $R^-$ defined by

\begin{eqnarray}
&& R^-|n \rangle = n(n-1), \quad n \geq 1 \\
&& R^-|0 \rangle = 0 \\
&& R^+|n \rangle = (n + 1) |n + 1 \rangle \label{fac1}
\end{eqnarray}
which obey the relations

\begin{equation}
R^+ = N\Gamma^+ = \Gamma^+ (N +1), \quad R^- = \Gamma^{-} N = (N+1) \Gamma^-
\end{equation}\\
We at once see that together with $N$, $R^+$ and $R^-$ the following algebra is obeyed

\begin{equation}
[R^-, N] = R^-, \quad [R^+, N] = -R^+, \quad [R^-, R^+] = 2N + 1
\end{equation}\\

Generically if there exist operators $\Omega_m^-$ and $\Omega_m^+$ defined by 

\begin{eqnarray}
&& \Omega_m^-|n \rangle = n(n-m), \quad n \geq m \\
&& \Omega_m^- |0 \rangle = 0, \quad n < m \\
&& \Omega_m^+ |n \rangle = (n + m) |n + m\rangle , \quad n = 0, 1, 2,... \label{fac1}
\end{eqnarray}
then these satisfy the relations

\begin{equation}
\Omega_m^- = (\Gamma^{-})^m N = (N+m) (\Gamma^-)^m, \quad \Omega_m^+ = N (\Gamma^{+})^m  = (\Gamma^+)^m (N+m) 
\end{equation}\\
where $m= 1, 2, ...$, then the following $su(1, 1)$ algebra is a straightforward consequence

\begin{equation}
[\Omega_m^-, N] = m\Omega_m, \quad [\Omega_m^+, N] = -m\Omega_m^+, \quad [\Omega_m^-, \Omega_m^+] = M(2N + m)
\end{equation}\\
It is evident that $(6.14)$ is a representative of the general feature $(6.19)$ for the $m = 1$ case.

By means of Bogoliubov transform one can disentangle the unitary operator 

\begin{equation}
U(\alpha) = e^{\alpha \Omega_m^+ - \alpha^* \Omega_m^-}, \quad  \alpha = \frac{r}{m} e^{i\phi}
\end{equation}\\
to express it in the form

\begin{equation}
U(\alpha) = e^{\frac{\beta}{m}e^{i\phi} \Omega_m^+} e^{\delta (\beta)} e^{-\frac{\beta}{m}e^{-i\phi} \Omega_m^-}
\end{equation}\\
where

\begin{equation}
\beta = \tanh r, \quad \delta (\beta) = \frac{2N + m}{2m} \ln (1 - \beta^2)
\end{equation}\\
As a result $U(\alpha)$ acting on the vacuum state $|0 \rangle$ turns out to be

\begin{equation}
U(\alpha) |0 \rangle = (1-\beta^2)^{\frac{1}{2}} \sum_{n=0}^\infty (\beta e^{i\phi})^n |mn \rangle, \quad m=1, 2, ...
\end{equation}\\

\section{HO squeezed states}

For the coherent state $(3.36)$ we saw that the argument of the exponential involved a linear complex combination of the annihilation and creation operators. From the point of view of representing a squeezed state one goes for the complex quadratic combination of these operators. The feature of squeezing corresponds to the possibility of arbitrary compression of the position uncertainty at the expense of appropriate fluctuation in the complementary momentum variable or vice-versa. The essential characteristic of a squeezed state in the HO potential is that although its profile is still Gaussian, its width is different from the ground state. 

In the next subsection we give a general framework to address single-mode squeezing. Subsequently we define a $\theta$-vacuum and make use of Bogoliubov transformation to connect with the usual vacuum state $|0\rangle $ of the Fock space. This allows us to provide a simple way to generate single-mode squeezing. Finally the phase connection is pointed out.

\subsection{Single-mode squeezing}

Squeezed states minimize the uncertainty relation but not restricting the ground state to belong to belong to this set. Comparing with $(3.47)$ the only difference is the appearence of the free value of the squeezing parameter or of the width $s$. Explicitly one writes in the coordinate representation for the normalized single-mode squeezed state the wave function

\begin{equation}
\psi_s (x) = (\pi s)^{-\frac{1}{4}} \exp \left [-\frac{1}{2s^2} (x-x_0)^2 + ip_0 x \right ], \quad \alpha \in \mathbb{C}
\end{equation}
where we identified $x_0 = \sqrt{2} \alpha_R$ and $p_0 = \sqrt{2} \alpha_I$ with $\alpha \in \mathbb{C}$ as before.

In terms of the operators $a$ and $a^\dagger$ the underlying squeezing operator is defined by 

\begin{equation}
S_\eta = e^{-\frac{\eta }{2}(a^2 -a^{\dagger 2})}, \quad \eta \in \R
\end{equation}\\
Employment of the $su(1, 1)$ algebra, highlighted in $(3.48) - (3.51)$, generated by a set of three operators $K^+ = \frac{1}{2} a^{\dagger^2}$, $K^- =  \frac{1}{2} a^2$ and $K^3 = -\frac{1}{2} [K^+ , K^-] = \frac{1}{2} (a^\dagger a + \frac{1}{2})$ and use of the entanglement formula $(6.21)$ make it convenient to recast $(7.2)$ in the following manner

\begin{equation}
S_\eta = e^{\frac{e^{i\phi}}{2} \tanh (r)a^{\dagger^2}} \left (\cosh (r) )^{-(a^\dagger a + \frac{1}{2})} \right ) e^{-{\frac{e^{-i\phi}}{2} \tanh (r)a^2}}
\end{equation}\\
The action of the operator $S_\eta$ on the vacuum $|0\rangle $ defines the squeezed state $|\eta\rangle $

\begin{equation}
|\eta \rangle = S_\eta |0 \rangle = \cosh^{-\frac{1}{2}}(r) \sum_{j=0}^\infty \left (\tanh (r) e^{i\phi}\right )^j \frac{\sqrt{(2j)!}}{j!} |2j \rangle
\end{equation}\\
The estimate of the mean photon number corresponds to $\langle a^\dagger a \rangle_\eta = \sinh^2 (r)$.

In the context of definitions of $R^+$ and $R^-$ presented in $(6.10) - (6.14)$. the representation in the form $\psi_R = e^{(\beta R^+ -\beta^* R^-)} |0 \rangle$ stands as  a squeezed state. So is $\psi_S = e^{(\gamma S^+ -\gamma^* S^-)} |0 \rangle$  \cite{Suk1, Suk2}, where the quantities $S^+$ and $S^-$ stand for $S^+ = N \Gamma^+2$ and $S^- = \Gamma^-2 N$ along with $\beta = \sigma e^{\xi}$, $\sigma = e^{i\xi}\tanh \sigma$,  $\delta = e^{i\tau \tanh \rho}$, $\gamma = \frac{\rho}{2} e^{i\tau}$.

\subsection{$\theta$-vacuum}

Let us apply Bogoliubov transformation \cite{Ume} on the annihilation and creation operators $a$ and $a^\dagger$ which generates $\theta$ dependent quantities $a_\theta$ and $a^\dagger_\theta$ through

\begin{equation}
a_\theta =S_\theta (a) S^{-1}_\theta, \qquad a^\dagger_\theta = S_\theta (a^\dagger) S^{-1}_\theta 
\end{equation}\\
where $a$ and $a^\dagger$ are related to $\hat{x}$ and $\hat{p}$ by means of $(2.10)$. Applying the BCH expansion $(46)$ the expressions for $a_\theta$ and $a^\dagger_\theta$ takes a 'hyperbolic' rotated form

\begin{equation}
\left[ \begin{array}{c} a_\theta  \\ a^\dagger_\theta \end{array} \right] = \begin{bmatrix} \cosh\theta  & -\sinh\theta \\ -\sinh\theta & \cosh\theta  \end{bmatrix}  \left[ \begin{array}{c} a \\ a^\dagger \end{array} \right]
\end{equation}\\

The vacuum $|0 \rangle$, which is annihilated by the operator $a$, moves over to the $\theta$-dependent entity $|0_\theta \rangle$ generated according to

\begin{equation}
 |0_\theta \rangle  = S_\theta|0 \rangle
\end{equation}\\
which is annihilated by $a_\theta$ 
\begin{equation}
a_\theta |0_\theta \rangle=0
\end{equation}\\
For the translation $a_\theta = a + \theta$, $|0_\theta \rangle$ obeys $a|0_\theta \rangle = -\theta |0_\theta \rangle$ and is therefore the coherent state with the eigenvalue $-\theta$.

Explicitly $ |0_\theta \rangle$ is given by

\begin{equation}
 |0_\theta \rangle  = e^{-\frac{1}{2}\ln\cosh\theta} e^{\frac{1}{2]}a^{\dagger 2}\tanh\theta}|0 \rangle
\end{equation}\\
which saturates the uncertainty product

\begin{equation}
 \langle (\Delta x_k)^2 \rangle  \langle (\Delta p_k)^2 \rangle = \frac{1}{4}
\end{equation}\\
In estimating $(7.10)$ we used the expectation value $\langle (\Delta x_k)^2 \rangle = \frac{1}{2} e^{2\theta}$ and $\langle (\Delta p_k)^2 \rangle = \frac{1}{2} e^{-2\theta}$with the help of the transformations $(2.11)$ and $(7.6)$ and the use of $(7.7)$, $(7.8)$ and $(7.9)$. Changing the sign of $\theta$, $\Delta x_k$ can be turned arbitrarily small by making $\theta$ appropriately large. This makes the idea of squeezing transparent.  

To demonstrate the validity of $(7.9)$ let us define the n-dependent vacuum value $u_n$ 

\begin{equation}
u_n = \langle 0|a^{2n}S_\theta|0 \rangle = \langle 0|a^{2n}e^{-\frac{\theta}{2}(a^2 -a^{\dagger 2})}|0 \rangle
\end{equation}\\
Differentiating with respect to $\theta$ shows

\begin{equation}
\frac{\partial}{\partial \theta} u_n = -\frac{1}{2} u_{n+1} + \frac{1}{2} \langle 0|a^{2n} a^{\dagger 2} e^{-\frac{\theta}{2}(a^2 -a^{\dagger 2})}|0\rangle 
\end{equation}\\

It is rather easy to work out the vacuum value in the right side by observing

\begin{equation}
 \frac{1}{2} \langle 0|a^2 a^{\dagger 2} = 1, \quad  \frac{1}{2} \langle 0|a^4 a^{\dagger 2} =6 a^2
\end{equation}\\
Employing them we are led to a recurrence relation 

\begin{equation}
\frac{\partial}{\partial \theta} u_n = -\frac{1}{2} u_{n+1} + n(2n -1) u_{n-1}
\end{equation}\\

It admits of the solution

\begin{equation}
u_n =k_n e^{-\frac{1}{2}\ln\cosh\theta} (\tanh \theta)^n
\end{equation}\\
where the pre-factor $k_n$ is given by 

\begin{equation}
k_n = \frac{(2n-1)!}{(n-1)!} \frac{1}{2^{n-1}}
\end{equation}\\

The case $n=1$ is relevant for our purpose for which $k_1=1$. It implies 

\begin{equation}
\langle 0|a^2|0_\theta \rangle = e^{-\frac{1}{2}\ln\cosh\theta} \tanh \theta
\end{equation}\\
Since $a^2{a^\dagger}^2=2+ 4a^\dagger a +(a^\dagger a)^2$, 
the solution $(7.9)$ of $|0_\theta \rangle$ follows. We note in passing that since $a_\theta|0_\theta \rangle =(a\cosh\theta -a^\dagger\sinh\theta)|0_\theta \rangle$, the annihilation character of the operator $a_\theta$ on the transformed vacuum state $|0_\theta \rangle$ becomes evident if we  expand the exponential $e^{\frac{1}{2}a^{\dagger 2} \tanh\theta}$. \\

\subsection{The phase connection}
 
As a final remark, the quantities $R^+$ and $R^-$ introduced in $(6.13)$, where the role of the phase operator $\Phi$ is implicit,
can also be used to generate the single-mode squeezed state $S_R$ \cite{Suk2}. Indeed corresponding to the form $(6.23)$ for $m = 1$ the squeezed state is given by 

 \begin{equation}
S_R  = e^{(\beta R^+ - \beta^* R^-)} |0 \rangle
\end{equation}\\
where notice that a linear combination of $R^{+}$ and $R^{-}$ appears in the argument of the exponential in contrast to the quadratic combination of $a^2$ and $a^{\dagger 2}$ in $(7.1)$. 

$S_R$ can also be projected as 

 \begin{equation}
S_R  = \left (1-|\beta|^2\right ) \sum_{n=0}^\infty \beta^n |n \rangle
\end{equation}\\
By $(6.20)$ and $(6.22)$ the quantities $\alpha$ and $\beta$ are

\begin{equation}
\alpha = r e^{i\phi}, \quad \beta = e^{i\nu} \tanh r 
\end{equation}\\
in which the phase factor $e^{i\nu}$ is introduced without any loss of generality.
 
 \section{Two-mode squeezing}
 
 In this section we discuss different ways of generating two-mode squeezing. First we lay out a general strategy of defining two sets of annijhilation and creation operators to act upon a two-mode vacuum state and attempt to generate the two-mode squeezed state related to the generators of the $SU(1, 1)$ group. In the next subsection, we again make use of Bogoliubov transformations to define the two-mode a squeezed  vacuum state. Finally, in section 9,  we propose a generalized quantum condition that makes clear how two-mode squeezing could be expressed as the joint product of two single mode squeezed states.

  \subsection{General framework}

Before realizing the two-mode squeezing state, let us focus on two decoupled single-mode operators $a_k$ and $a^\dagger_k$, $k=1,2$. In terms of coordinates and momenta these are similarly expressed as done in $(2.11)$
 
 \begin{equation}
x_k=\sqrt{\frac{\hbar}{2m\omega}}(a_k+a^\dagger_k), \quad p_k= -i\sqrt{\frac{m\hbar\omega}{2}}(a_k-a^\dagger_k), \quad k=1,2
\end{equation}
satisfying the quantum condition  

\begin{equation}
[x_k, p_k]=i\hbar, \quad k=1,2
\end{equation}

In terms of a parameter $s$ the squeezing operator $S$ is given by 

\begin{equation}
S  = e^{\frac{s}{2}(a_1a_2 - a_1^\dagger a_2^\dagger)} |0 \rangle
\end{equation}\\
To operate upon the two-mode vacuum state $|0, 0\rangle $ we define the $SU(1, 1)$ operators

\begin{equation}
K_- = a_1 a_2, \quad K_+ = a_1^\dagger a_2^\dagger, \quad K_0 = \frac{1}{2} \left (a_1 a_1^\dagger + a_2 a_2^\dagger -1\right )
\end{equation}\\
We use the general result \cite{Kli}

\begin{equation}
e^{\zeta_0 K_0 + \zeta_+ K_+ + \zeta_-K_-} = e^{\gamma_+ K_+} e^{\ln \gamma_0 K_0} e^{\gamma_- K_-}
\end{equation}\\
where $\zeta_0, \zeta_{\pm}$ are suitable real quantities and

\begin{equation}
\gamma_0 = \left(\cosh \theta - \frac{\zeta_0}{2\theta} \sinh \theta \right )^{-2}, \quad \gamma_{\pm} = \frac{2\zeta_{\pm} \sinh \theta}{2\theta \cosh \theta - \zeta_0 \sinh \theta}
\end{equation}\\
along with $\theta^2 = \frac{\zeta_0^2}{4} - \zeta_+\zeta_-$.

Putting $\zeta_0 =0$, $\zeta_{\pm} = \mp (\frac{s}{2})$, we see that $\theta$, $\gamma_0$ and $\gamma_{\pm}$ take the simple forms

\begin{equation}
\theta = \frac{s}{2}, \quad \gamma_0 = \left(\cosh (\frac{s}{2})\right)^{-2}, \quad  \gamma_{\pm} = \mp \tanh (\frac{s}{2})
\end{equation}\\
which imply 
\begin{equation}
e^{\frac{s}{2}(a_1a_2 - a_1^\dagger a_2^\dagger)} = e^{\tanh (\frac{s}{2}) a_1^\dagger a_2^\dagger} e^{-\ln  \cosh (\frac{s}{2})(a_1 a_1^\dagger + a_2 a_2^\dagger -1)} e^{\tanh (\frac{s}{2})a_1 a_2}
\end{equation}\\
In consequence the action of the two-mode squeezing operator on the vacuum reads

\begin{equation}
S |0, 0 \rangle  = \sqrt{1 - \lambda} \sum_{n = 0}^\infty |n, n \rangle, \quad \lambda = \tanh (s)
\end{equation}\\

In complete analogy to the above scheme two-mode squeezing comprising either the pair $(R^+, R^-)$ or $S^+, S^-)$ can be defined. We do not give the details here. \\

 \subsection{Bogoliubov transformations}
 
Applying the Bogoliubov matrix transformations on $a_k$ and $a^\dagger_k$, $k=1,2$ yield their $\theta$- dependent counterparts as in the earlier single-mode case

\begin{equation}
\left[ \begin{array}{c} a^\theta_k  \\ a^{\dagger\theta}_k \end{array} \right] = \begin{bmatrix} \cosh\theta  & -\sinh\theta \\ -\sinh\theta & \cosh\theta  \end{bmatrix}  \left[ \begin{array}{c} a_k \\ a^\dagger_k \end{array} \right]
\end{equation}\\

If $|0_\theta\rangle $ be the vacuum of $ a^\theta_k$, $k=1,2$, implying 

\begin{equation}
a^\theta_k |0_\theta \rangle =0
\end{equation}
then as before in $(7.5)$ we can solve for $|0_\theta\rangle $ to get

\begin{equation}
|0_\theta \rangle = e^{-\frac{1}{2}\ln\cosh\theta}e^{\frac{1}{2]}a_k^{\dagger 2}\tanh\theta}|0 \rangle
\end{equation}

Noting that we can easily calculate the uncertainties $\Delta x_1$ and $\Delta p_1$ by first transforming $x_1$ and $p_1$ in terms of $a^\theta_1 $ and $a^{\dagger\theta}_1$ by employing the two relations in $(8.1)$ and then using the transformation $(8.10)$. It turns out that 

\begin{equation}
\Delta x_1 = (\langle 0_\theta|x_1^2||0_\theta \rangle)^\frac{1}{2} =\sqrt{\frac{1}{2}} e^\theta, \quad \Delta p_1 = (\langle 0_\theta|p_1^2||0_\theta \rangle)^\frac{1}{2} =\sqrt{\frac{1}{2}} e^{-\theta}
\end{equation}
and similarly for $\Delta x_2$ and $\Delta p_2$ 

\begin{equation}
\Delta x_2 = (\langle 0_\theta|x_2^2||0_\theta \rangle)^\frac{1}{2} =\sqrt{\frac{1}{2}} e^\theta, \quad \Delta p_2 = (\langle 0_\theta|p_2^2||0_\theta \rangle)^\frac{1}{2} = \sqrt{\frac{1}{2}} e^{-\theta}
\end{equation}
It transpires that the two single-mode squeezed states saturate the uncertainty relation 

\begin{equation}
(\Delta x_k)^2(\Delta p_k)^2  = \frac{1}{4}, \quad k=1,2
\end{equation}
An interesting implication of either $(8.13)$ or $8.14)$ is that, due to the presence of the parameter $\theta$, we can make $(\Delta p_k)^2$, $k=1,2$ arbitrarily small by letting $\theta$ grow randomly large while for $(\Delta x_k)^2$, $k=1,2$  we can make it arbitrarily small by enforcing large negative values of $\theta$ and so have any amount of squeezing as one desires.\\

For two separate single-mode oscillators their coordinates and momenta are subject to the condition $(8.2)$. However for two-mode squeezing the basic commutations of $x_k$ and $p_k$ corresponding to the pair $(x_1, p_1)$ and $(x_2, p_2)$ need to be defined with opposite signs 

\begin{equation}
[x_1, p_1]=i\hbar, \quad [x_2, p_2]= -i\hbar, 
\end{equation}
This ensures that the Bogoliubov transformation $(7.6)$ is preserved. Note however that the sets of annihilation and creation operators $\gamma_1, \gamma_1^\dagger$ and $\gamma_2, \gamma_2^\dagger$ continue to hold the standard commutation relations 

\begin{eqnarray}
&& [a_1, a_1^\dagger]= 1, \quad [a_2, a_2^\dagger]= 1 \\ 
&& [a_1, a_2]= 0, \quad [a_1^\dagger, a_1]= 0
\end{eqnarray}

With this little background the two-mode squeezed vacuum state assumes the following representation

\begin{equation}
|0_\Theta \rangle = e^{-\frac{1}{2}\ln\cosh\Theta}e^{a_1^{\dagger}a_2^{\dagger}\tanh\Theta}|0 \rangle
\end{equation}
where the $\Theta$ denotes the squeezing parameter. It clearly reveals that in the modified vacuum $|0_\Theta \rangle$ the $a_1a_2$-pairs are condensed. Furthermore, the $a_1$ and $a_2$ modes contribute equally to $|0_\Theta \rangle$. It is also not difficult to establish the inequality 

\begin{equation}
(\Delta x_k)^2(\Delta p_k)^2  \geq \frac{1}{4} + \langle \Delta x_1 \Delta x_2 \rangle  \langle \Delta p_1 \Delta p_2 \rangle
\end{equation}
where the second factor in the right side amounts to $\frac{1}{4} \sinh^2 (2\Theta)$. It is the noise term which is typically prevalent in two-mode squeezing. 

\section{Generalized quantum condition}

In this section we derive pedagogically, in the framework of a generalized quantum condition, the result that a two-mode squeezed state can be written in a factorizable form with respect to the component single modes. We also examine the possibility of relating two-mode squeezing as a system of a pair of coupled oscillators.

To begin with let us combine the two commutations given by $(8.2)$ into a generalized quantum condition \cite{Bab}. For this we make use of the replacements

\begin{eqnarray}
&& p_1 \rightarrow -i \frac{\partial}{\partial x_1} -ig(x_2) \\ 
&& p_2 \rightarrow i \frac{\partial}{\partial x_2} -if(x_1)
\end{eqnarray}
implying that the two-mode squeezed state wave function $\psi_\Theta (x_1, x_2) = \langle x_1, x_2|0_\Theta \rangle$ would obey the differential equation

\begin{equation}
\left [ \mu\left (x_1 + \frac{\partial}{\partial x_1} +g(x_2) \right ) + \nu \left (x_2 - \frac{\partial}{\partial x_2} - f(x_1) \right ) \right ] \psi_\Theta (x_1, x_2) = 0
\end{equation}
where the quantities $\mu$ and $\nu$ are $\mu = \cosh \Theta$ and $\nu = - \sinh \Theta$.  Factorizing $\psi_\Theta (x_1, x_2)$ in the form $\phi (x_1) \chi (x_2)$ yields

\begin{eqnarray}
&& \mu\frac{d \phi}{dx_1} + [\mu x_1 - \nu f(x_1)]\phi = c\phi  \\ 
&& \nu \frac{d \chi}{dx_2} - [\nu x_2 + \mu g(x_2)]\phi = c\chi
\end{eqnarray}
where $c$ is a separation constant. Restricting $f(x_1) = -x_1$ and $g(x_2) = - x_2$ for normalizability of $\phi$ and $\chi$ leads to the solutions

\begin{eqnarray}
&& \phi = \phi_0 \exp \frac{1}{\mu} \left [cx_1 - (\mu + \nu) \frac{x_1^2}{2} \right ], \quad \phi_0 >0, \quad \mu + \nu > 0 \\ 
&&\chi = \chi_0 \exp \frac{1}{\nu} \left [cx_2 - (\mu - \nu) \frac{x_2^2}{2} \right ], \quad \chi_0 >0, \quad \mu - \nu < 0
\end{eqnarray}

With $z= \frac{1}{\sqrt{2}}(x_1 + i x_2)$ and $p_z= \frac{1}{\sqrt{2}}(p_1 + p_2)$ the basic commutation relations $(8.2)$ could be described by a single generalized quantum condition

\begin{equation}
[z, p_z] = i
\end{equation}
which admits of a plausible representation of of $p_z$

\begin{equation}
p_z = z - i\frac{\partial}{\partial z}
\end{equation}

It is interesting to note that if we introduce quantities $\Lambda_+$ and $\Lambda_-$ such that $\Lambda_{\pm} = \gamma_1 \pm i\gamma_2$, then $z$ and $p_z$ would read 

\begin{equation}
z = \frac{1}{2} (\Lambda_+ + \Lambda_-^\dagger), \quad p_z = \frac{i}{2} (\Lambda_+^\dagger - \Lambda_-)
\end{equation}
with the standard bosonic conditions $[\Lambda_{\pm}, \Lambda_{\pm}^\dagger] = 1$ holding. Furthermore, since $\Lambda_{\pm} |0 \rangle = (\gamma_1 \pm i\gamma_2)|0,0 \rangle = 0$, $\Lambda_{\pm}$ have the same ground states as those of $\gamma_1$ and $\gamma_2$. Moreover, because $\gamma_1 = \frac{1}{2} (\Lambda_+ + \Lambda_-)$ and $\gamma_2 = \frac{1}{2i} (\Lambda_+ - \Lambda_-)$, the two-mode squeezed state given by

\begin{equation}
|0_\Theta \rangle = e^{-\ln \cosh \Theta} e^{\gamma_1^\dagger \gamma_2^\dagger \tanh \Theta}|0,0 \rangle
\end{equation}
is also expressible as 

\begin{equation}
|0_\Theta \rangle = e^{-\ln \cosh \Theta} e^{\frac{i}{2} (\Lambda_+\dagger^2 - \Lambda_-\dagger^2) \tanh \Theta} |0,0 \rangle
\end{equation}
The above result makes it transparent that squeezing for two-mode quantum systems \cite{Arv} can be interpreted \cite{Mil, Bab} as the direct product of two single-mode squeezed states. Even squeezed multi-mode wave function can be shown to arise \cite{Caves1, Caves2} from the single-mode squeezed state for the normal modes. \\

\section{SQM approach}

To deal with the construction of coherent states for a wider class of potentials, other than the HO, some works have already been reported \cite{Mmn5, Bha, Shree, Ger, Djf1, Nan, Mol}. Here we  summarize the answer to the specific question that if two potentials are isospectral to each other (in other words, they have the same eigenvalues and the S-matrix) then how their coherent states and squeezed states are related\footnote{For some comments related to the isospectral issue see \cite{Rosu}.} In this regard, it is useful to follow the approach of \cite{Kha} which, by adopting the techniques of SQM\footnote{ For reviews on $SQM$ see \cite{Cks1, Gj, Bag1, Cks2, Rod, Djf2} and references therein. The key feature of SQM rests on the factorization method \cite{Inf, Don, Lev, And} (sometimes
also expressed through intertwining relationships) that allows one to uncover insightful properties of the underlying system.}, has related the coherent states for the strictly isospectral Hamiltonians by a unitary transformation. Let us remark that in one dimensional case, strict isospectrality
could be present in the first-order systems when the factorization energy E is smaller
than the ground-state of the starting Hamiltonian. In such a context, the potential and eigenfunctions of the partner Hamiltonian of the SQM system are known
in terms of analytic expressions, which are, in general, more complex than the corresponding ones of the initial Hamiltonian. 

Continuing with the units of $\hbar = \omega = m = 1$, let us define operators 

\begin{equation}
A = \frac{1}{\sqrt{2}} \left ( \frac{d}{dx} + W(x) \right ), \quad
A^\dagger = \frac{1}{\sqrt{2}} \left ( -\frac{d}{dx} + W(x) \right )
\end{equation}\\
where $W(x)$ is the so-called superpotential and is an arbitrary function of $x$. These operators go over to the corresponding $a$ and $a^\dagger$ in $(2.10)$ for the particular choice of $W(x) = 1$ which conforms to the HO case. In the language of $SQM$ the Hamiltonians formed by the combinations $A^\dagger A$ and $AA^\dagger$ are the partner pairs. This has the implication that their eigenvalues, eigenfunctions and $S$-matrices are related. Put another way, if one can solve for the eigenfunctions of $A^\dagger A$ then one at once has a knowledge of $AA^\dagger$.

Factorization of a quantum Hamiltonian is however not unique \cite{Mie}. To this end if we consider a new set of operators 

\begin{equation}
B = \frac{1}{\sqrt{2}} \left ( \frac{d}{dx} + \hat{W(x)} \right ), \quad
B^\dagger = \frac{1}{\sqrt{2}} \left ( -\frac{d}{dx} + \hat{W(x)} \right )
\end{equation}\\
subject to $BB^\dagger = AA^\dagger$ then this results into a Riccati equation whose solution is provided by the relation 

\begin{equation}
\hat{W} (x) = W (x) + \phi_\lambda (x)
\end{equation}\\
where $\phi_\lambda (x)$ stands for 

\begin{equation}
\phi_\lambda (x) = \psi_0^2 (x) \left [ \lambda + \int_{-\infty}^x  \psi_0^2 (y)dy \right ]^{-1}
\end{equation}\\
where $\lambda \in \Re$ but outside of $[-1,0]$ and $\psi_0 (x)$ is the normalized ground state wave function corresponding $A^\dagger A$. 

Taking $E_0$ to be the lowest energy eigenvalue, it is not difficult to establish that the eigenstates $\chi_n$ of the strictly isospectral family of Hamiltonians $H_\lambda = B^\dagger B + E_0$ are linked to the eigenstates $\psi_n$ of $H = A^\dagger A + E_0$ from the equality $(B\dagger B) b^\dagger A = b^\dagger A (A^\dagger A)$. As a result one has the following normalized expressions 

\begin{eqnarray}
&&\chi_0 (x) = \sqrt{\lambda (\lambda + 1)} \left [ \lambda + \int_{-\infty}^x  \psi_0^2 (y)dy \right ]^{-1} \psi_0 (x)\\
&&\chi_n (x) = \psi_n (x) + \frac{1}{2(E_n - E_0)} \phi_\lambda (x) \left ( \frac{d}{dx} + W(x) \right ) \psi_n (x), \quad n = 1, 2,...
\end{eqnarray}\\

That a unitary transformation induced by the operator $U$ relates $H$ and $H_\lambda$ can be shown by noticing that if one writes $B = a U^\dagger$ then it is evident that $U^\dagger U = 1$. With a new operator defined by $\tilde{a} = Ua U^\dagger$ implying $B^\dagger B = \tilde{a}^\dagger \tilde{a} = U A^\dagger A U^\dagger$ the following result immediately follows

\begin{equation}
H_\lambda = U H U^\dagger
\end{equation}\\

The unitary character of $U$ emerges i.e. 

\begin{equation}
U^\dagger U =1
\end{equation}
by observing that 
$E_n = \langle \chi_n|H_\lambda|\chi_n\rangle $, from which it follows that $U^\dagger |\chi_n\rangle  = |\psi_n\rangle $, and the orthonormality of the sets $|\psi_n\rangle $ and $|\chi_n\rangle $, namely, $\langle \psi_n|\psi_m\rangle  = \langle \chi_n|UU^\dagger|\chi_m\rangle  = \delta_{nm}$.

$(9.7)$ and $(9.8)$ are the central results that serve as a good benchmark to determine the coherent states. Suppose that the Hamiltonian is expressible as a linear combination of the generators $J_i, i = 1,2,3$ of a Lie group $G$ obeying the algebra\footnote{For the determination of coherent states induced by nonlinear algebras see \cite{Sunil}.}

\begin{equation}
H =  \sum_i d_i J_i, \quad d_i \in \mathbb{C}
\end{equation}
where the algebra is given by the commutation relations

\begin{equation}
[J_i, J_j] = c_{ij}^k J_k, \quad i = 1,2,3
\end{equation}
Then one can write 

\begin{equation}
H_\lambda =  \sum_i d_i \tilde{J}_i, \quad \tilde{J_i} = U J_i U^\dagger
\end{equation}
$H_\lambda$ being unitarily related to $H$. 

It therefore follows that for an element $D(z)$ belonging to the coset space of $G$, the coherent states corresponding to the isospectral Hamiltonians $H$ and $H_\lambda$ are of the types

\begin{eqnarray}
&& |z \rangle = D(z) |\psi_0 \rangle, \\
&& |z; \lambda \rangle = D_\lambda |\chi_0 \rangle
\end{eqnarray}
where $D_\lambda = U D(z) U^\dagger$. The above two expressions can be identified as the Perelomov coherent states \cite{Per} associated with $H$ and $H_\lambda$ respectively.

For the particular case of the oscillator Hamiltonian $H$ given by $(2.13)$, the associated $H_\lambda$ can be identified to be 

\begin{equation}
H_\lambda = \tilde{a}^\dagger \tilde{a} + \frac{1}{2}, \quad \tilde{a} = UaU^\dagger, \quad \tilde{a}^\dagger = U a^\dagger U^\dagger
\end{equation}
where $[\tilde{a}, \tilde{a}^\dagger] = 1$, the coherent states associated withe isospectral oscillator family $|z; \lambda\rangle $ are given by

\begin{equation}
\tilde{a} |z; \lambda \rangle = z |z; \lambda \rangle
\end{equation}
where the eigenvalue $z$ is independent of $\lambda$. In the Perelomov sense the following holds

\begin{equation}
|z ; \lambda \rangle = D_\lambda (z) |\chi_0 \rangle, \quad D_\lambda (z) = e^{z\tilde{a}^\dagger - z^* \tilde{a}}
\end{equation}
satisfying the minimum uncertainty product. 

Even squeezed states of isospectral Hamiltonians are related by unitary transformation. As explicitly shown in \cite{Kha} one can also construct a more general state which minimizes the uncertainty product in analogy with the squeezed coherent state of the HO in terms of the usual displacement operator 

\begin{equation}
|\xi, z;\lambda \rangle = S_\lambda (\xi)D_\lambda (z) |\chi_0 \rangle, \quad S_\lambda (\xi) = e^{\frac{1}{2} (\xi \tilde{a}\dagger^2 - \xi^* \tilde{a}^2)}
\end{equation}
Indeed it turns out that the individual position and momentum uncertainties are unequal while the product assumes the value of one-half in accordance with minimum value of the uncertainty principle. Further, the state $|\xi, z; \lambda \rangle$ is connected to the squeezed coherent
state of the $HO$ by a unitary transformation.

Sometimes obtaining commuting sets of
creation and annihilation operators \cite{Fuk} can be effectively used to construct coherent states. For the complex parity-time symmetric oscillator\footnote{Exploration of complex non-Hermitian quantum systems, in particular, those  admitting a combined parity and time reflection symmetry, is an area of active research interest and a rapidly evolving field (see, for example,
the comprehensive treatments in \cite{CMB1}). We might recall that the interest in this area of research stems from the conjecture, made nearly two decades ago \cite{BB}, that the eigenvalues
of the governing Hamiltonian would normally support a real bound-state spectrum unless a spontaneous breaking of the underlying parity-time-symmetry takes place. In the latter case the accompanying energies cease to be real and as a consequence eigenvalues in conjugate
pairs develop signaling a phase transition in the system.} \cite{Zno}, a pair of commuting sets of creation and annihilation operators was derived \cite{Bagchi} that allowed building of coherent states as eigenstates of such annihilation operators. In such a pursuit a modified normalization
integral \cite{Bag3} was employed, as appropriate for parity-time systems, although it turned out that the coherent states were only normalizable in the open interval (0, 1) for the guiding coupling parameter. 

Construction of new families of coherent states has also been reported for other non-Hermitian systems \cite{Broy} using the  Gazeau-Klauder approach \cite{Gkl}. (See also \cite{Tri}.) In particular 
coherent states of the parity-symmetric Scarf I potential has been determined.

\newpage

\section{Summary}

To summarize, in this review, we delved into that specialized branch of quantum mechanics which concerns with certain non-classical aspects and which comes under the category of what are termed as coherent states and squeezed states. For coherent states, we focused on three different strategies of constructing them  namely, the annihilation operator or ladder operator approach, displacement-operator approach and minimum-uncertainty approach. In all the three cases, we tried to give a complete coverage by pointing out their inter-connectivity and highlighting some of the special properties. In this way we tried to cover the basic knowledge of the field. For squeezed states, after laying out a general strategy, we considered in some detail different aspects of the generation of both single-mode and two-mode squeezing and also emphasized on the feasibility of Bogoliubov transformations. We also derived a generalized quantum condition which demonstrated how the factorization principle works to express multi-mode squeezing, specifically the  two-mode squeezing, as a direct product of single-mode states.

\section{Acknowledgments}

AK would like to thank the Indian National Science Academy (INSA) for the award of INSA Senior Scientist 
position. We also thank Prasanta K Panigrahi and Barry C Sanders for informative ommunications.


\newpage


















\begin{thebibliography}{99}

\bibitem{Sch} E. Sch\"rodinger, Der stetige ¨ubergang von der mikro-zur
makromechanik,  Naturwissenschaften. {\bf 14}, 664 (1926).\\

\bibitem{Mmn1} M. M. Nieto and L. M. Simmons Jr., Coherent States for General Potentials, Phys. Rev. Lett. {\bf 41}, 207 (1978).\\
 
\bibitem{Mmn2} M. M. Nieto and L. M. Simmons Jr., Coherent states for general potentials. I. Formalism, Phys. Rev. {\bf D20}, 1321 (1979). \\

\bibitem{Mmn3} M. M. Nieto, L. M. Simmons Jr. and V. P. Gutschick, Coherent states for general potentials. VI. Conclusions about the classical motion and the WKB approximation, Phys. Rev. {\bf D23}, 927 (1981). \\

\bibitem{Walls} D. F. Walls and G. J. Milburn,  Quantum Optics, Springer (Science and Business Media), 2007.\\

\bibitem{Kla1} J. R. Klauder, Continuous‐Representation Theory. I. Postulates of Continuous‐Representation Theory, J. Math. Phys. {\bf 4}, 1055 (1963). \\

\bibitem{Kla2} J. R. Klauder and E. C. G. Sudarshan, Fundamentals of Quantum Optics, W. A. Benjamin, New York, 1968.\\

\bibitem{Gla} R. J. Glauber, Photon Correlations, Phys. Rev. Lett. {\bf 10}, 84 (1963).\\

\bibitem{Sud} E. C. G. Sudarshan, Equivalence of Semiclassical and Quantum Mechanical Descriptions of Statistical Light Beams, Phys. Rev. Lett. {\bf 10}, 277 (1963).\\

\bibitem{Mmn4} M. M. Nieto, Coherent States and Squeezed States,
Supercoherent States and Supersqueezed States, arXiv:9212116 (hep-th).\\

\bibitem{Gaz1} J.P. Gazeau, Coherent states in Quantum Optics: An oriented overview in Integrability, Supersymmetry and Coherent States, Springer (2019).\\


\bibitem{Fer} A. Ferraro, S. Olivares, M. G. A. Paris, Gaussian states in continuous variable quantum information, Lecture notes, Bibliopolis, Napoli, 2005.\\


\bibitem{Barut} A. O. Barut and L. Girandello, New “Coherent” states associated with non-compact groups, Commun. Math. Phys. {\bf 21}, 41 (1971). \\

\bibitem{Radc} J. M. Radcliffe, Some properties of coherent spin states, J. Phys. (Math. Gen.) {\bf A4}, 313 (1971). \\

\bibitem{Per} A. Perelomov, Generalized Coherent States and Their Applications, Springer, Berlin, 1986. \\

\bibitem{Bor} V. V. Borzov and E. V. Damaskinsky,  Generalized Coherent States for the q-Oscillator Associated with Discrete q-Hermite Polynomials. J. Math. Sc. {\bf 132-1}, 26 (2006).  \\

\bibitem{Com} M. Combescure and D. Robert, Coherent States and Applications in Mathematical Physics. Springer, 2012. \\

\bibitem{Que1} C. Quesne, New q-deformed coherent states with an explicitly known resolution of unity, J. Phys. (Math. Gen.) {\bf A 35}, 9213 (2002). \\

\bibitem{Que2} C. Quesne, Generalized coherent states associated with the $C_\lambda$-extended Oscillator,  Ann. Phys. {\bf 293}, 147 (2001). \\

\bibitem{Ruby} V. Chithiika Ruby and M. Senthilvelan, On the construction of coherent states of position-dependent mass Schr\"{o}dinger equation endowed with effective potential,  J. Math. Phys. {\bf 51}, 052106 (2010). \\

\bibitem{Vyas} V. M. Vyas, Airy wavepackets are coherent states,  Am. J. Phys. {\bf 86}, 750 (2018). \\

\bibitem{Ber} M. V. Berry and N. L. Balazs, Nonspreading wave packets, Am. J. Phys. {\bf 47}, 264 (1979). \\

\bibitem{Gaz2} J. P. Gazeau, Coherent states in Quantum Information: An example of experimental manipulations, J. Phys. (Conf. Series) {\bf 213}, 012013 (2010).\\

\bibitem{San} B. C. Sanders, Review of Entangled Coherent States,
J. Phys. (Math. and Theor) {\bf A45}, 244002 (2012).\\ 	

\bibitem{Buk} S. H. Bukhari, S. Aslam, F. Mustafa, A. Jamil, S. N. Khan and M. A. Ahmad, Entangled coherent states for quantum information processing, Optik {\bf 125} (2014) 3788. \\ 

\bibitem{Zha} W-M. Zhang, Coherent states in field theory, arXiv:9908117 (hep-th).\\

\bibitem{Ver} M. Verschuren, Coherent states in quantum mechanics, Bachelor thesis, Radbond University, Nijemgen (2011).\\

\bibitem{Hae} T. M. van Haeringen, Generalized coherent states, Bachelor thesis, University of Groningen (2016).\\

\bibitem{Dey}   S. Dey, A. Fring and V. Hussin, A squeezed review on coherent states and nonclassicality for non-Hermitian systems with minimal length, Springer Proc. Phys. {\bf 205}, 209 (2018). \\

\bibitem{Ort}  O. Rosas-Ortiz, Coherent and squeezed states: introductory review of basic notions, properties and generalizations, submitted as a contribution to the monograph "Integrability, Supersymmetry and Coherent states", in honour of Prof. V. Hussin, Eds., S. Kuru, J. Negro and L. M. Nieto (2019), arXiv:1812.07523 (quant-ph).\\

\bibitem{Berg} H. Bergerson and J.-P. Gazeau, Integral quantizations with two basic examples, Ann. Phys., {\bf 344}, 43 (2014).\\

\bibitem{Ali} S. T. Ali, J. P. Antoine and J.-P. Gazeau, Coherent states, Wavelets and their generalizations, Theor. Math. Phys., Springer,   2000 (Chapter 11).\\

\bibitem{Fre} L. Freidel and E. R. Livine, U(N) Coherent States for Loop Quantum Gravity, arXiv: 1005.2090 (gr-qc).\\

\bibitem{Hus1} D. J. Fernandez C, V. Hussin and  O. Rosas-Ortiz, Coherent states for Hamiltonians generated by supersymmetry, J. Phys. (Math. Gen.) {\bf A40}, 6491 (2007).\\

\bibitem{Fring} B. Bagchi and A. Fring, Quantum, noncommutative and MOND corrections to the entropic law of gravitation, Int. J. Mod. Phys. {\bf B33}, 1950018 (2019).\\

\bibitem{Bagchi} B. Bagchi and C. Quesne, Creation and annihilation operators and coherent states for the PT-symmetric oscillator, Mod. Phys. Lett. {\bf A16}, 2449 (2001).\\

\bibitem{Hus2} S. Dey, V. Hussin, Noncommutative q-photon added coherent states, Phys. Rev. {\bf A 93}, 053824 (2016).\\

\bibitem{Che} V. Y. Chernyak, S. Choi and S. Mukamel, Generalized coherent state representation of Bose-Einstein condensates, Phys. Rev. {\bf A 67}, 053604 (2003).\\


\bibitem{Skg} J Klauder and B Skagerstam, Coherent states: Applications in physics and mathematical physics, World Scientific (1985). \\

\bibitem{Spe} Special issue on coherent states: mathematical and physical aspect,  Guest Eds. S. T. Ali, J. P. Antoine, F. Bagarello and J.-P. Gazeau, J. Phys. (Math. Theor.) {\bf A 44}, 260201 (2011).\\

\bibitem{Gsa} G. S. Agarwal, Nonclassical statistics of fields in pair coherent states, J. Opt. Soc. Am. {\bf B5} (1988) 1940.\\

\bibitem{Dim} M. T. DiMario, L. Kunz, K. Banaszek and F. E. Becerra, Optimized communication strategies with binary coherent states over phase noise channels, npj Quantum Inf {\bf 5}, 65 (2019).\\

\bibitem{Hol} J. N. Hollenhorst, Oscillations in photon distribution of squeezed states,  Phys. Rev.  {\bf D19}, 1669 (1979). \\

\bibitem{Lou} R. Loudon, The Quantum Theory of Light, Oxford University Press, 2000. \\

\bibitem{Sinha} S. Sinha, J. Emerson, N. Boulant,
E. M. Fortunato, T. F. Havel and D. G. Cory, Experimental Simulation of Spin Squeezing
by Nuclear Magnetic Resonance, Quantum Information Processing, {\bf 2}, 433 (2003). \\

\bibitem{Gsr} G. S. Agarwal and R. Simon, A new representation for squeezed states, Opt. Comm. {\bf 92}, 105 (1992). \\

\bibitem{Koro} A. N. Korotov, Notes on squeezed states in x-space representation, arXiv:1809.05531 (quant-ph). \\

\bibitem{Whe} W. Schleich and J. A. Wheeler, Oscillations in photon distribution of squeezed states,  J. Opt. Soc. Am. {\bf 4}, 1715 (1987). \\

\bibitem{Eke} A. K. Ekert and P. L. Knight, Correlations and squeezing of two-mode oscillators, Am. J. Phys. {\bf 57}, 692 (1989).\\

\bibitem{Bars}  L. Barsotti, J. Harms and R. Schnabel, Squeezed vacuum states of light for gravitational wave detectors,  Rep. of Prog. in Phys. {\bf 82}, 016905 (2019).\\

\bibitem{Lvo} A. I. Lvovsky, Squeezed light, Section in book: Photonics Volume 1: Fundamentals of Photonics and Physics, pp. 121 - 164, Ed. D. Andrews, Wiley, UK, 2015.\\
\bibitem{Rsc} R. Schnabel, Squeezed states of light
and their applications in laser interferometers, 	Phys. Rep. {\bf 684}, 1 (2017). \\

\bibitem{Bar} S. M. Barnett, General criterion for squeezing, Opt. Commun.{\bf 61}, 432 (1998). \\

\bibitem{Roy} B. Roy, Nonclassical properties of the even and odd intermediate number squeezed states, Phys. Lett.{\bf B12}, 23 (1998). \\

\bibitem{Boy} R. W. Boyd, Nonlinear Optics, Academic Press, 2003.\\

\bibitem{Wal} H. Walther, B. T. H. Varcoe, B-G. Englert and T. Becker, Cavity quantum electrodynamics, Rep. Prog. Phys. {\bf 69}, 1325 (2006).\\

\bibitem{Pag} D. Pagnoux et al, Photonic Crystals: Towards Nanoscale Photonic Devices, 2008.\\


\bibitem{Mply} L. Inzunza, M. S. Plyushchay and  A. Wipf, Conformal bridge between freedom and confinement, Phys. Rev. {D 101} 105019 (2020) . \\

\bibitem{Aba} J. Abadie et al. (The LIGO Scientific Collaboration), A gravitational wave observatory operating beyond the quantum shot-noise limit, Nat. Phys. {\bf 7}, 962 (2011). \\ 

\bibitem{Aas} J. Aasi et al., Enhanced sensitivity of the LIGO gravitational wave detector by using squeezed states of light, Nat. Photon. {\bf 7}, 613 (2013).\\

\bibitem{Lug} L. A. Lugiato, A. Gatti and  E. Brambilla, Quantum imaging, Quantum Imaging, J. Opt. B: Quantum and Semiclassical Optics. {\bf 4} (2002) S176. \\

\bibitem{Gat} A. Gatti, T. Corti and E. Brambilla, Squeezing and EPR correlation in the Mirrorless Optical Parametric Oscillator, arXiv:1704.09010 (quant-ph)\\

\bibitem{Jma} J. Ma, X. Wang, C. P. Sun and  F. Noria, Quantum spin squeezing, Phys. Rep. {\bf 509}, 89 (2011).\\ 


\bibitem{Arv} Arvind, B. Dutta, N. Mukunda and R. Simon,  Two mode quantum systems: invariant classification of squeezing transformations and squeezed states, Phys. Rev. {\bf A52}, 1609 (1995).\\

\bibitem{VG} A. Venugopalan and R. Ghosh,  Wigner-function description of quantum-mechanical nonlocality, Phys. Rev. {\bf A 44}, 6109 (1991).\\

\bibitem{Kir} A. Kireev, A. Mann, M. Revzen and H. Umezawa, Thermal squeezed states in thermo field dynamics and quantum and thermal fluctuations, Phys. Lett. {\bf A 142}, 215 (1989). \\

\bibitem{Cha} S. Chaturvedi, R. Sandhya, V. Srinivasan and R. Simon, Thermal counterparts of nonclassical states in quantum optics, Phys. Rev. {\bf A41}, 3969 (2000).\\

\bibitem{Wig} E. Wigner, On the quantum correction for thermodynamic equilibrium, Phys. Rev. {\bf 40}, 749 (1932).\\

\bibitem{Mil}  G. J. Milburn, Multimode minimum uncertainty squeezed states, J. Phys. (Math. Gen.) {\bf A 17}, 737 (1984).\\

\bibitem{Bab} B. Bagchi and D. Bhaumik, A reappraisal of two-mode squeezing and intermode coupling, Mod. Phys. Lett. 
{\bf A15}, 825 (2000); see also,  B. Bagchi and D. Bhaumik,  Coherent states for paraboson, J. Phys. (Math. Gen.) {\bf A 30}, L593 (1997). \\

\bibitem{Aga} G. S. Agarwal, Generation of Pair Coherent States and Squeezing via the Competition of Four-Wave Mixing and Amplified Spontaneous Emission, Phys. Rev.Lett.
{\bf 57}, 827 (1986).

\bibitem{Gho1} R. Ghosh and G. S. Agarwal, Theory of a two-photon squeezed laserlike oscillator, Phys. Rev. 
{\bf A39}, 1582 (R) (1989).  \\

\bibitem{AG} A. Agarwal and R. Ghosh, Two-photon squeezed laser with long-lived atoms, Phys. Rev. 
{\bf A50}, 1950 (1994).  \\

\bibitem{Ade} G. Adesso, S. Ragy and A. R. Lee, Continuous variable quantum information: Gaussian states and
beyond, Open Systems \& Information Dynamics, {\bf 21}, World Scientific, 2014\\

\bibitem{Brau} S. L. Braunstein and P. Van Loock, Quantum information with continuous variables, Rev. Mod. Phys., {\bf 77}, 2 (2005). \\

\bibitem{Gasp} M. Gasperini and M. Giovannini, Quantum squeezing and cosmological entropy production, Classical and Quantum Gravity {\bf 10}, 9 (1993).\\

\bibitem{Schn} R. Schnabel, Squeezed states of light and their applications in laser interferometers, Phys. Rep. {\bf 684}, 1
(2017).\\

\bibitem{Gar} A. Garcia-Chung, Squeeze operator: a classical view, arXiv:2003.04257 (math-ph). \\

\bibitem{Gho2} R. Ghosh and L. Mandel, Observation of nonclassical effects in the interference of two photons, Phys. Rev. Lett. {\bf 59}, 1903 (1987). \\

\bibitem{GHOM} R. Ghosh, C. K. Hong, Z. Y. Ou and L. Mandel, Interferenced of two photons in parametric down conversion,  Phys. Rev. {\bf A 34}, 3962 (1986). \\

\bibitem{Scu}  M. O. Scully and M.S. Zubairy, Quantum Optics, Cambridge University Press, 1997. \\

\bibitem{Tei} M. C. Teich and B. A. E. Saleh, Quantum Optics {\bf 1}, 151 (1989).\\

\bibitem{Kla3} J. R. Klauder and B-S. Skasgerstam Coherent states: Applications in Physics and Mathematical Physics, World Scientific Publishing Company, 1985.\\

\bibitem{Dir1} P. A. M. Dirac, The Principles of Quantum Mechanics, Oxford University Press, 1999.\\

\bibitem{Gas} S. Gasiorowicz, Quantum Physics, Wiley, New York, 1974.\\

\bibitem{Bra} B.H. Bransden and C.J. Joachain, Quantum Mechanics, Pearson, 2000.\\

\bibitem{Sak} J.J. Sakurai, Modern Quantum Mechanics, Addison-Wesley Publishing Company, Inc., 1994.\\

\bibitem{Acd} A. C. Davies, Quantum Mechanics, DAMTP Lecture notes, 2015.\\

\bibitem{Kla4} J. R. Klauder, Continuous‐Representation Theory. II. Generalized Relation between Quantum and Classical Dynamics, J. Math. Phys. {\bf 4}, 1058 (1963).\\

\bibitem{Iwa} G. Iwata, Non-Hermitian operators and eigenfunction expansions, Prog. Theor. Phys. {\bf 6}, 216 (1951).\\

\bibitem{Dut} D. Bhaumik, T. Nag and B. Dutta-Roy, Coherent states for angular momentum, J. Phys. (Math. Gen.) {\bf A 8}, 1868 (1975). \\

\bibitem{Band} A. Bandyopadhyay, Coherent states for angular momentum, Ph. D. Thesis (Physics), Indian Institute of Technology, Kanpur, 1996. \\

\bibitem{Oje} D. Ojeda-Guillen, R. D. Mota and V. D. Granados, The $SU(1, 1)$ Perelomov number coherent states and the
non-degenerate parametric amplifier, J. Math. Phys. {\bf 55}, 042109 (2014). \\

\bibitem{Gerr}  C. C. Gerry, Dynamics of SU (1, 1) coherent states, Phys. Rev. {\bf A 31}, 2721 (1985). \\

\bibitem{Gsb} G. S. Agarwal and A. Biswas, Quantitative measures of entanglement in pair
coherent states, J. Opt. {\bf  B7}, 350 (2005).\\ 

\bibitem{Gag} A. Gabris and  G. S. Agarwal,  Quantum teleportation with pair-coherent states, Int. J. Quant. Inf {\bf 5}, 17 (2007).\\

\bibitem{Wu} R-K. Wu, S-B. Li, Q-M. Wang and J-B. Xu,  Application of pair coherent states in quantum information processes, Mod. Phys. Lett. {\bf 17}, 913 (2004).\\

\bibitem{Gil}  A. Gilchrist, P. Deuar and M. D. Reid, Contradiction of quantum mechanics with local hidden variables for quadrature phase measurements on pair-coherent states and squeezed macroscopic superpositions of coherent states, Phys. Rev. {\bf A 60}, 4259 (1999).\\

\bibitem{Tara} K. Tara  and G. S. Agarwal, Einstein-Podolsky-Rosen paradox for continuous variables using radiation fields in the pair-coherent state,  Phys. Rev. {\bf A50}, 2870 (1994).\\

\bibitem{Has} W. S. Chung and H. Hassanabadi,  Generalized pair coherent states and non-classical properties, Eur. Phys. J. Plus {\bf 134}, 394 (2019). \\

\bibitem{Bbd} D. Bhaumik, K. Bhaumik and B. Dutta-Roy, Charged bosons and the coherent state,  J. Phys.(Math. and Gen.) {\bf A9}, 1507 (1976). \\

\bibitem{Wang1}X-G. Wang, Two-mode nonlinear coherent states, arxiv:0004042 (quant-ph).\\

\bibitem{Gst} G. S. Agarwal and K. Tara, Nonclassical properties of states generated by the excitations on a coherent state,  Phys. Rev. 
{\bf A43}, 492 (1991).\\

\bibitem{Siva} S. Sivakumar, Photon added coherent states as nonlinear coherent states, J. Phys.(Math. and Gen.) {\bf A32}, 3411 (1999).\\

\bibitem{Wang2} X-G. Wang, B. C. Sanders and S-H. Pan, Entangled coherent states for systems with SU(2) and SU(1,1) symmetries, J. Phys. (Math. and Gen.) {\bf A33}, 7451 (2000).\\ 

\bibitem{Sha} P. Shanta, S. Chaturvedi, V. Srinivasan, G. S. Agarwal, and C. L. Mehta,
 Unified approach to multiphoton coherent states, Phys. Rev. Lett. {\bf 72}, 1447 (1994).\\ 



\bibitem{Spi} V. Spiridonov, Universal superpositions of coherent states and self-similar potentials, Phys. Rev. 
{\bf A52}, 1909 (1995).\\

\bibitem{Yur} B. Yurke and D. Stoler, Generating quantum mechanical superpositions of macroscopically distinguishable states via amplitude dispersion,
Phys.Rev. Lett. {\bf 57}, 13 (1986). \\

\bibitem{Gkl} J.-P. Gazeau and J. R. Klauder, Coherent states for systems with discrete and continuous spectrum, J. Phys. (Math. and Gen.) {\bf A32}, 123 (1999).\\


\bibitem{Don} S.H. Dong, Factorization Method in Quantum Mechanics, Springer, The Netherlands
(2007).\\

\bibitem{Lev} G. Levai, A search for shape-invariant solvable potentials, J. Phys.(Math. and Gen.) {\bf A22}, 689 (1989).\\

\bibitem{And} A. A.Andrianov, M. V. Ioffe and V. P. Spiridonov, Higher-Derivative Supersymmetry and the Witten Index, Phys. Lett. {\bf A174}, 273 (1993).\\

\bibitem{Mie} B. Mielnik, Factorization Method and New Potentials with the
Oscillator Spectrum, J. Math. Phys {\bf 25}, 3387 (1984).\\

\bibitem{Sunil} V. Sunilkumar, B. A. Bambah, R. Jagannathan, P. K. Panigrahi and V. Srinivasan, Coherent states of nonlinear algebras: applications to quantum optics, J. Opt. (Quant. Semicl.) {\bf B2}, 126 (2000).\\

\bibitem{Dod} V. V. Dodonov, 'Nonclassical' states in quantum optics: a 'squeezed' review of the first 75 years, J. Opt. B: Quantum Semiclass. Opt. {\bf 4}, R1 (2002).\\

\bibitem{Hei} W. Heitler, Quantum theory of Radiation, Oxford University Press, Oxford, 1954 \\

\bibitem{Dir2} P. A. M. Dirac, Quantum theory of Emission and Absorption in Quantum Electrodynamics, Dover Publication, New York, 1958 \\

\bibitem{Sus} L. Susskind and J. Glogower, Quantum mechanical phase and time operator, Physics {\bf 1}, 49 (1964).\\ 

\bibitem{Car1} P. Carruthers and M. M. Nieto, Coherent states and the number-phase uncertainty relation, Phys. Rev. Lett.
{\bf 14}, 387 (1965).\\

\bibitem{Car2} P. Carruthers and M.M. Nieto, Phase and Angle Variables in Quantum Mechanics, Rev. Mod. Phys. {\bf 40}, 411 (1968).\\

\bibitem{Kli} A. B.  Klimov and S. M.  Chumakov, A group-theoretical approach to quantum optics:  models of atom-field interactions, Wiley-VCH, 2009.\\

\bibitem{Suk1} C. V. Sukumar, Revival Hamiltonians, phase operators and non-Gaussian squeezed states, J. Mod. Opt. {\bf 36}, 1591 (1989).\\

\bibitem{Suk2} C. V. Sukumar, A non-Gaussian single-mode squeezed state of the simple harmonic oscillator, J. Phys. (Math. Gen.) {\bf A 21}, L1065 (1988). \\

\bibitem{Ume} H. Umezawa, Advanced Field Theory, AIP, 1993.\\

\bibitem{Caves1} C. M. Caves and B. L. Schumaker, New formalism for two-photon quantum optics. I. Quadrature phases and squeezed states, Phys. Rev. 
{\bf A31}, 3068 (1985).\\

\bibitem{Caves2} B. L. Schumaker and C. M. Caves, New formalism for two-photon quantum optics. II. Mathematical foundation and compact notation, Phys. Rev. 
{\bf A31}, 3093 (1985).\\

\bibitem{Mmn5} M. M. Nieto and L. M. Simmons Jr., Coherent states for general potentials. II. Confining one-dimensional examples, Phys. Rev. {\bf D20}, 1332 (1979).\\

\bibitem{Shree} T. Shreecharan, P. K. Panigrahi, and J. Banerji, Coherent states for exactly solvable potentials, Phys. Rev. {\bf 69}, 012102 (2004).\\

\bibitem{Bha} D. Bhaumik, B. Dutta-Roy and G. Ghosh, Classical limit of the hydrogen atom, J. Phys. (Math. Gen.) {\bf A 19}, 1355 (1986). \\

\bibitem{Ger} C. C. Gerry, Coherent states and the Kepler-Coulomb problem,  Phys. Rev. {\bf A33}, 6 (1986). \\

\bibitem{Djf1} D. J. Fernandez C, V. Hussin and M. L. Nieto, Coherent states for isospectral oscillator Hamiltonians, J. Phys. (Math. Gen.) {\bf A 27}, 3547 (1994). \\

\bibitem{Nan} S. Nandi and C. S. Sastry, Classical limit of the two-dimensional and the three-dimensional hydrogen atom, J. Phys. (Math. Gen.) {\bf A 22}. 1005 (1989). \\

\bibitem{Mol} Marcin Molski, A general scheme for the construction of minimum uncertainty coherent states of anharmonis oscillators, J. Phys. (Math. Theor.) {\bf A 42}. 165301 (2009). \\

\bibitem{Kha} M. S. Kumar and A. Khare, Coherent States for Isospectral Hamiltonians, Phys. Lett. {\bf A217}, 73 (1996).\\

\bibitem{Rosu} H. C. Rosu, Comment on quant-ph/9509008 by Kumar and Khare, arXiv:quant-ph/9608017\\

\bibitem{Cks1} F. Cooper, A. Khare and U. Sukhatme, Supersymmetry and quantum mechanics, Phys.Rep. {\bf 251}, 267 (1995) and references therein.\\

\bibitem{Gj} G. Junker, Supersymmetric Methods in Quantum and Statistical Physics
(Springer, 1996).\\

\bibitem{Bag1} B. Bagchi, Supersymmetry in Quantum and Classical Mechanics (Chapman
and Hall/CRC, Florida, 2000).\\

\bibitem{Cks2} F. Cooper, A. Khare and U. Sukhatme, Supersymmetry in quantum mechanics, World Scientific (2001). \\

\bibitem{Rod} R. de Lima Rodrigues, The quantum mechanics SUSY algebra: An introductory review, arXiv:hep-th/0205017 (2002).\\

\bibitem{Djf2} D. J. Fernandez, Trends in supersymmetric quantum mechanics, submitted as a contribution to the monograph "Integrability, Supersymmetry and Coherent states", in honour of Prof. V. Hussin, Eds., S. Kuru, J. Negro and L. M. Nieto (2019), arXiv:1811.06449.\\

\bibitem{Inf} L. Infeld and T.E. Hull, The factorization method, Rev.Mod.Phys. {\bf 23}, 21 (1951).\\

\bibitem{Fuk} T. Fukui and N. Aizawa, Shape-invariant potentials and an associated coherent state, Phys. Lett. {\bf A180}, 308 (1993). \\

\bibitem{CMB1}  C. M. Bender, PT Symmetry: In Quantum and Classical Physics (World Scientific, 2019).\\

\bibitem{BB}  C. M. Bender and S. Boettcher, Real Spectra in Non-Hermitian Hamiltonians Having PT Symmetry, Phys. Rev. Lett. {\bf 80}, 5243 (1998). \\

\bibitem{Zno} M. Znojil, PT-symmetric harmonic oscillators, Phys. Lett. {\bf A259}, 220 (1999).\\

\bibitem{Bag3} B. Bagchi, C. Quesne and M. Znojil, Generalized continuity equation and modified normalization in PT-Symmetric Quantum Mechanics,  Mod. Phys. Lett. {\bf A16}, 2047 (2001).\\

\bibitem{Broy} B. Roy and P. Roy, Coherent states of non-Hermitian quantum systems, Phys. Lett. {\bf A359}, 110 (2006).\\

\bibitem{Tri} D. A. Trifonov, Pseudo-Boson coherent and Fock  states, Differential Geometry, Complex Analysis and Mathematical Physics, eds. K. Sekigawa et al, pp. 241-250, World Sc., 2009.









\end{thebibliography}
\end{document}